\documentclass[10pt,twocolumn,letterpaper]{article}
\usepackage[utf8]{inputenc}
\usepackage[T1]{fontenc}
\usepackage{mathptmx}
\usepackage[scaled=0.90]{helvet}
\usepackage{amssymb,amsmath}
\usepackage[letterpaper,textwidth=7in,textheight=9in,centering,columnsep=0.3in,headsep=14pt,headheight=12pt,footskip=26pt]{geometry}
\usepackage[compact]{titlesec}
\titleformat{\section}{\normalfont\large\bfseries}{\thesection}{0.6em}{}
\titleformat{\subsection}{\normalfont\normalsize\bfseries}{\thesubsection}{0.6em}{}
\titleformat{\subsubsection}{\normalfont\normalsize\bfseries\itshape}{\thesubsubsection}{0.6em}{}
\titlespacing*{\section}{0pt}{10pt plus 3pt minus 2pt}{5pt plus 1pt}
\titlespacing*{\subsection}{0pt}{8pt plus 2pt minus 2pt}{3pt plus 1pt}
\titlespacing*{\paragraph}{0pt}{5pt plus 2pt minus 1pt}{0.7em}
\usepackage{microtype}

\DisableLigatures[-]{encoding = T1, family = tt*}
\usepackage{graphicx}
\usepackage{booktabs}
\usepackage{multirow}
\usepackage{array}
\usepackage{tabularx}
\newcolumntype{Y}{>{\raggedright\arraybackslash}X}
\newcolumntype{P}[1]{>{\raggedright\arraybackslash}p{#1}}
\usepackage{makecell}
\usepackage{longtable}
\usepackage[table]{xcolor}
\usepackage{tikz}
\usetikzlibrary{arrows.meta,positioning,shapes.geometric,shapes.misc,fit,backgrounds,calc,decorations.pathreplacing,shadows.blur,matrix}
\usepackage{pgfplots}
\pgfplotsset{compat=1.17}
\usepgfplotslibrary{groupplots}
\usepackage[ruled,vlined,linesnumbered]{algorithm2e}
\usepackage{listings}
\usepackage[font=small,labelfont=bf]{caption}
\usepackage{subcaption}
\usepackage{enumitem}
\usepackage[most]{tcolorbox}
\usepackage{pifont}
\usepackage{fancyhdr}
\usepackage[numbers,sort&compress]{natbib}
\usepackage{xurl}
\usepackage[hidelinks]{hyperref}
\usepackage{cleveref}
\usepackage{placeins}
\usepackage{needspace}
\usepackage{float}

\definecolor{tkorange}{HTML}{EB6834}
\definecolor{tkblue}{HTML}{2A78D6}
\definecolor{tkaqua}{HTML}{1BAF7A}
\definecolor{tkink}{HTML}{2B2A28}
\definecolor{tkgrey}{HTML}{6B6A66}
\definecolor{tkred}{HTML}{C0392B}
\definecolor{tkpurple}{HTML}{7A4FD0}
\definecolor{tkfill}{HTML}{F5F4F1}
\definecolor{tkcream}{HTML}{FFF4EC}
\definecolor{tkbluefill}{HTML}{EAF2FC}
\definecolor{tkaquafill}{HTML}{E6F6EF}
\definecolor{tkredfill}{HTML}{FBEAE8}
\definecolor{tkpurplefill}{HTML}{F1ECFB}

\newcommand{\code}[1]{\texttt{#1}}
\newcommand{\yes}{\textcolor{tkaqua}{\ding{51}}}
\newcommand{\no}{\textcolor{tkred}{\ding{55}}}

\newtcolorbox{claimbox}[1][]{enhanced,colback=tkcream,colframe=tkorange,boxrule=0.6pt,arc=2pt,
  left=6pt,right=6pt,top=4pt,bottom=4pt,fonttitle=\bfseries\small,coltitle=white,#1}
\newtcolorbox{limitbox}[1][]{enhanced,colback=tkredfill,colframe=tkred!80,boxrule=0.6pt,arc=2pt,
  left=6pt,right=6pt,top=4pt,bottom=4pt,fonttitle=\bfseries\small,coltitle=white,#1}

\setlist{itemsep=1.5pt,topsep=3pt}

\newcommand{\nohyph}{\hyphenpenalty=10000\exhyphenpenalty=10000\relax}
\renewenvironment{abstract}{\section*{\centering Abstract}\noindent\ignorespaces}{\par\medskip}

\newcommand{\papertitle}{Tracekit Box: Running AI Agents in Micro-VMs\\Where Every Way Out Is Decided and Signed}
\fancypagestyle{first}{\fancyhf{}\fancyfoot[C]{\small\thepage}}
\hypersetup{pdftitle={Tracekit Box: Running AI Agents in Micro-VMs Where Every Way Out Is Decided and Signed},pdfauthor={Bravish Ghosh}}

\begin{document}
\twocolumn[{%
  \centering
  \vspace*{-0.35in}
  {\LARGE\bfseries \papertitle\par}
  \vspace{12pt}
  {\large Bravish Ghosh\par}
  \vspace{3pt}
  {\normalsize Cygnux Labs \quad\textperiodcentered\quad \texttt{info@cygnuxlabs.com}\par}
  \vspace{2pt}
  {\normalsize Code (MIT): \url{https://github.com/Cygnux-Labs/Tracekit-box}\par}
  \vspace{12pt}
\begin{minipage}{\textwidth}
\centering
\resizebox{\linewidth}{!}{%
\begin{tikzpicture}[x=1cm,y=1cm,font=\sffamily\scriptsize,
  box/.style={draw=tkink!70,rounded corners=2pt,fill=white,align=center,inner sep=2.5pt,line width=0.5pt,blur shadow={shadow blur steps=4,shadow xshift=0.4pt,shadow yshift=-0.6pt,shadow opacity=22}},
  door/.style={draw=tkorange!90!black,rounded corners=2pt,fill=tkcream,align=left,inner sep=3pt,line width=0.6pt,text width=4.1cm},
  zone/.style={rounded corners=5pt,line width=0.8pt,dashed},
  zlabel/.style={font=\sffamily\bfseries\scriptsize,inner sep=3pt,anchor=north west},
  ar/.style={-{Stealth[length=1.9mm]},line width=0.6pt,draw=tkink},
  arb/.style={{Stealth[length=2.2mm]}-{Stealth[length=2.2mm]},line width=1.1pt,draw=tkink},
  lab/.style={font=\sffamily\scriptsize,fill=white,inner sep=1pt,text=tkink},
]
\fill[tkredfill!70,rounded corners=5pt] (0,0) rectangle (4.6,10.2);
\draw[zone,tkred!75] (0,0) rectangle (4.6,10.2);
\node[zlabel,text=tkred!80!black] at (0,10.2) {MICRO-VM \textperiodcentered{} untrusted};

\fill[tkcream,rounded corners=5pt] (5.3,0) rectangle (12.0,10.2);
\draw[zone,tkorange] (5.3,0) rectangle (12.0,10.2);
\node[zlabel,text=tkorange!80!black] at (5.3,10.2) {HOST \textperiodcentered{} trusted: \texttt{tkbox daemon}};

\fill[tkaquafill,rounded corners=5pt] (13.2,5.0) rectangle (17.8,10.2);
\draw[zone,tkaqua!80!black] (13.2,5.0) rectangle (17.8,10.2);
\node[zlabel,text=tkaqua!60!black] at (13.2,10.2) {SIGNER \textperiodcentered{} its own trust domain};

\fill[tkbluefill,rounded corners=5pt] (13.2,2.55) rectangle (17.8,4.6);
\draw[zone,tkblue!80] (13.2,2.55) rectangle (17.8,4.6);
\node[zlabel,text=tkblue!80!black] at (13.2,4.6) {INTERNET};

\fill[tkpurplefill!70,rounded corners=5pt] (13.2,0) rectangle (17.8,2.15);
\draw[zone,tkpurple!80] (13.2,0) rectangle (17.8,2.15);
\node[zlabel,text=tkpurple] at (13.2,2.15) {ANYONE, OFFLINE};

\node[box,minimum width=4.1cm,text width=3.9cm,fill=tkredfill!30] (agent) at (2.3,8.55) {\textbf{The agent}\\Claude Code, Codex or any program, its tools and its model calls};
\node[box,minimum width=4.1cm,text width=3.9cm] (guest) at (2.3,6.75) {\textbf{\texttt{tkbox-guest}} (PID 1)\\exec \textperiodcentered{} files \textperiodcentered{} diff \textperiodcentered{} boot config};
\node[box,minimum width=4.1cm,text width=3.9cm] (fwd) at (2.3,5.15) {\textbf{egress forwarder}\\\texttt{127.0.0.1:3128} (\texttt{HTTP\_PROXY})};
\node[box,minimum width=4.1cm,text width=3.9cm] (ws) at (2.3,3.5) {\textbf{\texttt{/workspace}}\\a copy of your checkout, no \texttt{.git}, no credential files};
\node[box,minimum width=4.1cm,text width=3.9cm,fill=tkredfill] (nic) at (2.3,1.6) {\textbf{no network card}\\no DNS, no UDP; secrets only as placeholders \texttt{tkbox\_GH\_3f9c\ldots}};

\node[door] (dcmd) at (8.65,9.05) {\textbf{Commands}: each \texttt{tkbox exec} decided before it starts};
\node[door] (dnet) at (8.65,7.55) {\textbf{Network}: allowlist; others wait for a person; private, metadata and own addresses refused};
\node[door] (dsec) at (8.65,6.0) {\textbf{Secrets}: real value swapped into one header, only for bound hosts; leaks blocked};
\node[door] (dmod) at (8.65,4.6) {\textbf{Model calls}: egress to a model host, recorded};
\node[door] (dchg) at (8.65,3.15) {\textbf{Changes}: patch at stop; \texttt{tkbox apply} decides each file before writing};
\node[door] (dstop) at (8.65,1.75) {\textbf{Stop}: timeout, \texttt{tkbox stop}, or a signer hold pauses or kills the VM};
\node[box,text width=4.1cm,fill=tkfill] (you) at (8.65,0.55) {\textbf{Person on the host}: \texttt{tkbox approve} \textperiodcentered{} your checkout \textperiodcentered{} your keys};

\node[box,minimum width=4.2cm,text width=4.0cm,fill=tkaquafill!40] (signer) at (15.5,8.3) {\textbf{Tracekit signer}\\holds the key; decides with its own policy; signs each call into a per-run hash chain inside a Merkle tree};
\node[box,minimum width=4.2cm,text width=4.0cm] (run) at (15.5,6.0) {\textbf{one run per box}\\\texttt{Bash}, \texttt{WebFetch}, \texttt{Write} decisions; \texttt{tkbox.*} observations; close reason};

\node[box,minimum width=4.2cm,text width=4.0cm] (net) at (15.5,3.5) {allowlisted hosts \textperiodcentered{} model APIs\\\texttt{api.github.com} gets the real key};
\node[box,minimum width=4.2cm,text width=4.0cm,fill=white] (tkb) at (15.5,1.05) {\textbf{\texttt{run.tkb}} + \texttt{run.trust.json}\\\texttt{tracekit verify} $\to$ \textcolor{tkaqua!70!black}{\textbf{VERIFIED}}};

\draw[ar] (agent) -- (guest);
\draw[ar] (agent.west) -- ++(-0.15,0) |- (fwd.west);
\draw[arb,tkink] (guest.east) -- node[lab,above,pos=0.42]{vsock 1024} (dcmd.west);
\draw[ar,line width=1.1pt] (fwd.east) -- node[lab,above,pos=0.42]{vsock 1025} (dnet.west);
\draw[line width=0.6pt] (6.3,7.55) -- (6.3,4.6);
\draw[ar] (6.3,6.0) -- (dsec.west);
\draw[ar] (6.3,4.6) -- (dmod.west);
\draw[ar] (ws.east) -- node[lab,above,pos=0.4]{patch} (dchg.west);
\foreach \d in {dcmd,dnet,dsec,dmod,dchg,dstop} \draw[ar,tkaqua!60!black] (\d.east) -- (12.5,0 |- \d.east);
\draw[line width=0.9pt,tkaqua!60!black] (12.5,9.05) -- (12.5,1.75);
\draw[ar,tkaqua!60!black,line width=0.9pt] (12.5,8.3) -- (signer.west);
\node[font=\sffamily\scriptsize\bfseries,text=tkaqua!50!black,align=center] at (12.6,9.75) {decide\\\& sign};
\draw[ar,tkaqua!60!black] (signer) -- (run);
\draw[ar,tkblue,line width=0.7pt,preaction={draw=tkcream,line width=3pt,-}] (dnet.east) ++(0,-0.3) -- (12.9,7.25) -- (12.9,3.5) -- (net.west);
\draw[ar,tkpurple,line width=0.7pt] (run.east) -- (18.15,6.0) -- node[lab,fill=white,text=tkpurple,rotate=90,pos=0.5]{\texttt{tkbox export}} (18.15,1.05) -- (tkb.east);
\draw[ar,tkaqua!60!black] (you.north) -- (dstop.south);
\end{tikzpicture}}

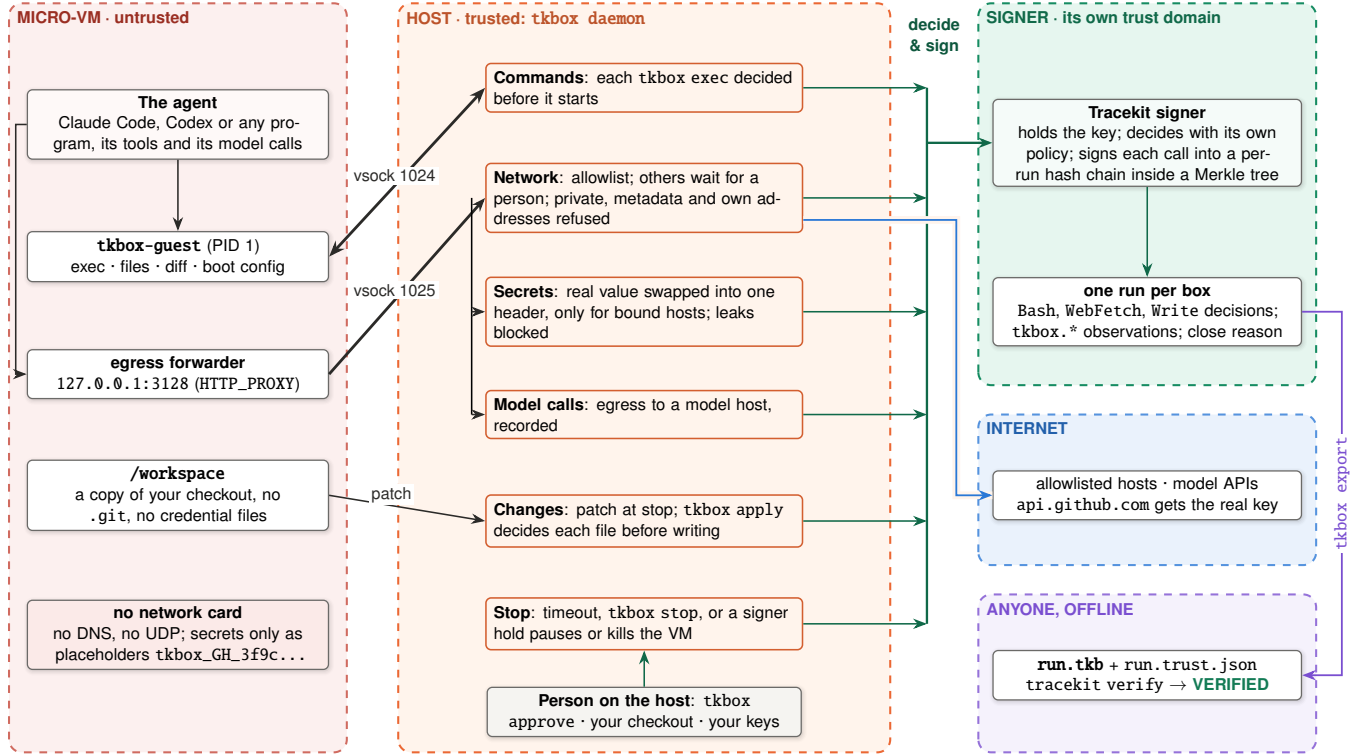
\captionof{figure}{\textbf{The architecture of Tracekit Box.} Each agent job runs in its own Firecracker micro-VM with no network card. The only ways out are vsock ports served by the daemon on the host: commands, the network (and through it secrets and model calls), and the workspace patch. At each of these doors the daemon asks a Tracekit signer, which the VM cannot reach, to decide and sign what passes. Real keys stay on the host and are added on the way out, only for the hosts they are bound to. When the job ends, the box's record exports as one \texttt{.tkb} bundle that anyone can verify offline.}
\label{fig:arch}
\end{minipage}

  \vspace{10pt}
}]
\thispagestyle{first}

\begin{abstract}
\noindent AI agents now do real jobs: they install packages, call model APIs with their operators' keys, and change code in repositories that deploy to production. To do that they need a network, credentials and a way to hand back their work, and each of these is also a way to leak data, steal a key or change a system nobody approved. Sandboxes contain the agent but leave no evidence of what crossed their boundary; tracing records what happened but contains nothing. We present Tracekit Box, an open-source system (MIT licence, v0) that does both. Each job runs in its own Firecracker micro-VM with no network card. The only ways out are a few doors on the host, served over vsock: commands started in the box, outbound connections, secrets, model calls and the workspace patch. At every door a Tracekit signer the agent cannot reach decides what passes and signs it into one run per box. Names are resolved on the host, so the box makes no DNS queries; private, link-local and cloud-metadata addresses are refused whatever an allowlist says; secrets never enter the VM, which sees only placeholders that the host proxy swaps for real values in one header on requests to the secret's bound hosts; and the agent's changes come back as a patch that is applied to the checkout only file by file, after the signer has decided each file. Everything fails closed: without a signer there is no daemon, and a record Box must keep but cannot stops the box. Each job ends as a single \code{.tkb} bundle that anyone can verify offline with \code{tracekit verify}.

We report what the v0 implementation has been tested against, and what it has not. All 13 end-to-end checks pass on a real KVM runner in continuous integration, including a box with no network interface, refusal of the cloud metadata address, a secret swapped in for its bound host and blocked in another header, and an exported bundle that verifies; the test that runs them, including image pull and boot, takes 28.5\,s. Fifteen fuzz targets cover the proxy's address parsing, TLS ClientHello checks, secret redaction, the guest protocol and the patch parser, and produced nine findings, all fixed. The demo run in the repository holds 102 signed records (46 decisions) and verifies. Box does not yet run Firecracker's jailer, supports one host with a co-located signer, and records the commands an agent starts inside the VM only where they cross a door.
\end{abstract}

\section{Introduction}\label{sec:intro}

Coding agents such as Claude Code and Codex, and agents built on frameworks such as the OpenAI Agents SDK or LangGraph, are moving from suggesting changes to making them. A typical job looks small: fix a failing test, upgrade a dependency, open a pull request. To do it the agent needs three things that are also the three ways it can do harm. It needs the \textbf{network}, to install packages and call its model, and the network is how data leaves. It needs \textbf{credentials}, a model API key and often a GitHub token, and a credential the agent can read is a credential the agent can send anywhere. And it needs a way to hand back its \textbf{work}, which is a way to change files nobody reviewed, including the CI workflows that run on the next push.

The text an agent reads (a README, an issue, a web page, a tool result) is often written by someone else, so an agent can be steered into misusing all three by indirect prompt injection~\cite{greshake2023indirect,zhan2024injecagent,debenedetti2024agentdojo}. It can also misuse them without any attacker: in September 2026 OpenAI reported that a research agent in a training sandbox reached an outside chatbot by tunnelling through DNS, a path its network filter did not cover~\cite{openai_dns}.

Two families of tools address this today, and each covers half of it.
\begin{itemize}[leftmargin=*]
  \item \textbf{Sandboxes} contain the agent: containers, gVisor~\cite{young2019gvisor}, micro-VMs such as Firecracker~\cite{agache2020firecracker}, and agent-specific sandboxes~\cite{e2b,srt,openshell,microsandbox}. They decide what may leave, but they leave little evidence of what did, and the evidence they keep is a log on the same machine, written by the same operator.
  \item \textbf{Recording} tools capture what an agent did. Tracekit~\cite{tracekit} goes furthest: every tool call is decided and signed by a signer the agent cannot reach, and a run exports as an offline-verifiable bundle. But a signer can sign only what reaches it. An agent in an ordinary environment can open a socket, read a key from its environment or write a file without passing a hook.
\end{itemize}

Tracekit Box joins the two. It puts the agent in a room whose only exits are doors the host controls, and puts a Tracekit signer at every door. The design follows from one rule: \emph{nothing leaves the VM except through a door, and nothing passes a door without a signed decision}.

\begin{claimbox}[title={The claim this paper defends}]
For a job run in Tracekit Box with a signer the agent cannot reach: every command started in the box from the host, every outbound connection and every file written to the checkout by \code{tkbox apply} was decided by the signer before it happened, and is in a signed record that anyone can verify offline; a bound secret's value never enters the VM and is sent only in its bound header to its bound hosts. Box does \emph{not} record what the agent does inside the VM without crossing a door, and does not judge what an allowed host is sent. A verified bundle proves the record is intact, not that the job was correct or benign.
\end{claimbox}

\paragraph{Contributions.}
\begin{enumerate}[leftmargin=*]
  \item \textbf{A door model for agent sandboxes} (\cref{sec:threat,sec:design}): a micro-VM with no network card whose only exits are vsock ports served by the host, with each exit (commands, network, secrets, model calls, the workspace patch, stop) mediated and decided.
  \item \textbf{Egress for agents without DNS} (\cref{sec:network}): names resolved once on the host and dialled by checked address, refusal of private, link-local, metadata and the host's own addresses in every form, SNI bound to the requested host, and a person in the loop for hosts off the allowlist.
  \item \textbf{Secrets the agent never holds} (\cref{sec:secrets}): placeholders inside the VM, swapped for the real value in one header for exact bound hosts, with leak blocking and response redaction on every connection the proxy reads.
  \item \textbf{Changes that come back as decided writes} (\cref{sec:changes}): a strict patch pipeline in which \code{git apply} never runs on the checkout, git must read the patch exactly as Box does, sensitive files need explicit consent, and every file is decided before anything is written.
  \item \textbf{Signed, fail-closed records} (\cref{sec:record}): one Tracekit run per box, with a mapping from each door to signed decisions and observations, and no silent loss of a record Box must keep.
  \item \textbf{An open implementation and its tests} (\cref{sec:impl,sec:eval}): about 25,600 lines of Go with 340 unit tests, 15 fuzz targets, an end-to-end suite on real KVM (two test functions, 13 checks), a GitHub Action, an MCP server and a Python SDK.
\end{enumerate}

\paragraph{Relation to Tracekit.} Tracekit Box is a separate project. It uses an unmodified Tracekit 1.4.2 signer through its public RPC and Tracekit's evidence format, and adds what Tracekit alone cannot provide: an environment in which the signer's view is complete for the doors Box controls, because the agent has no other way out. This paper describes Box; Tracekit's signer, evidence format and verifier are described in~\cite{tracekit}.

\paragraph{Roadmap.} \Cref{sec:related} compares Box with sandboxes and recorders. \Cref{sec:threat} gives the threat model. \Cref{sec:design} describes the box and its doors, and \cref{sec:record} how they are recorded. \Cref{sec:impl} covers the implementation and its interfaces, \cref{sec:eval} the evaluation, and \cref{sec:discussion} the limits and future work.

\section{Background and related work}\label{sec:related}

\paragraph{Isolation.} Containers share the host kernel, so a kernel bug in one container is a bug in all of them. gVisor intercepts system calls with a user-space kernel, at a measurable cost~\cite{young2019gvisor}. Kata Containers run each container in a lightweight VM~\cite{kata}. Firecracker is a minimal virtual machine monitor built for serverless workloads: a small device model, a few emulated devices, and boot times in the low hundreds of milliseconds~\cite{agache2020firecracker}. Box uses Firecracker because a hardware-virtualisation boundary with a small device model is the strongest isolation that still starts fast enough to give every job its own VM, and because Firecracker's vsock device lets the host offer the guest exactly the channels it chooses and nothing else.

\paragraph{Agent sandboxes.} Several systems now run agents in isolated environments. E2B offers cloud sandboxes for agents built on Firecracker~\cite{e2b}. microsandbox runs self-hosted micro-VMs for untrusted code~\cite{microsandbox}. Anthropic's sandbox-runtime confines a local agent with operating-system sandboxing and filters its network through a proxy by domain~\cite{srt}, and NVIDIA OpenShell runs agents in sandboxes whose outbound requests pass a policy~\cite{openshell}. These systems decide what an agent may reach. As far as we know none of them, as shipped, produces a record of those decisions that a third party can verify, keeps the agent's credentials out of the sandbox while still letting it use them, or returns the agent's changes through a decided, per-file write path.

\paragraph{Egress control.} Allowlisting by domain is the common way to limit where an agent may connect. Its known weak points are the ones an agent can find: DNS itself as a channel~\cite{openai_dns}, names that resolve to private or metadata addresses (server-side request forgery), DNS rebinding between the check and the connect, and domain fronting, where the TLS server name and the HTTP host differ. Box closes each of these at the door (\cref{sec:network}) rather than in the guest, because the guest is the attacker.

\paragraph{Credential brokering.} Keeping a secret out of the process that uses it is an old idea: credential helpers, cloud workload identity, and proxies that add authentication headers on the way out. Box applies it to agents: the VM holds only a placeholder, and the host proxy adds the real value for exact hosts in one header, blocks the placeholder or the value anywhere else on traffic it reads, and redacts the value from responses (\cref{sec:secrets}).

\paragraph{Tamper-evident records of agent actions.} Tracekit~\cite{tracekit} records every tool call through a signer the agent cannot reach, in a per-run hash chain inside an RFC~6962 Merkle tree~\cite{rfc6962} whose checkpoints witnesses can cosign, and exports runs as offline-verifiable bundles. Its limit is completeness: an action that never reaches the signer leaves no record. Box supplies the missing half for the actions it mediates: in a VM with no other exit, a connection, a secret use or a file write that leaves the box must pass a door, and every door asks the signer.

\begin{table*}[t]
\centering\small
\caption{What each approach provides for an agent job. \yes{} provided, $\circ$ partial, -- not provided. Rows describe the approach as typically deployed.}\label{tab:related}
\footnotesize
\begin{tabular}{@{}P{4.6cm}cccccc@{}}
\toprule
Approach & \makecell{Hardware\\isolation} & \makecell{Egress\\decided} & \makecell{Key never\\in sandbox} & \makecell{Changes\\decided per file} & \makecell{Signed,\\offline-verifiable} & \makecell{Fails\\closed} \\
\midrule
Container with a network policy & -- & $\circ$ & -- & -- & -- & $\circ$ \\
gVisor / Kata & $\circ$\,/\,\yes & $\circ$ & -- & -- & -- & $\circ$ \\
Agent sandboxes (E2B, microsandbox, OpenShell, sandbox-runtime) & $\circ$ & \yes & -- & -- & -- & $\circ$ \\
Tracing and observability & -- & -- & -- & -- & -- & -- \\
Tracekit alone & -- & $\circ$ & -- & $\circ$ & \yes & \yes \\
\midrule
\textbf{Tracekit Box (v0)} & \yes & \yes & \yes & \yes & \yes & \yes \\
\bottomrule
\end{tabular}
\end{table*}

\Cref{tab:related} summarises the comparison. The point of Box is not any one column: each has been done before. It is that the columns hold together, so that the record is complete for the doors, and the doors are the only ways out.

\section{Threat model}\label{sec:threat}

\subsection{Trust domains}

Box has three trust domains (\cref{tab:domains}). The attacker controls everything inside the VM: the agent, everything it runs, the guest kernel's view of the world, and every byte the guest agent sends. We assume the agent is hostile, either prompt-injected or following instructions from a file it read. The attacker does not control the host, the signer, or the person who approves connections on the host.

\begin{table}[h]
\centering\footnotesize
\caption{Trust domains.}\label{tab:domains}
\begin{tabularx}{\linewidth}{@{}P{1.45cm}P{1.6cm}Y@{}}
\toprule
Domain & Trust & What runs there \\
\midrule
Host & trusted & \code{tkbox daemon}, the egress proxy, the image builder, the secrets passed in, the user's checkout \\
Guest (micro-VM) & untrusted & the agent, everything it runs, and the guest agent \code{tkbox-guest} itself \\
Signer & trusted, separate from the agent & the Tracekit 1.4.2 signer and its signing key, on the same host as the daemon, reached only over a Unix socket \\
\bottomrule
\end{tabularx}
\end{table}

\subsection{What the attacker wants, and what stops it}

\Cref{tab:goals} lists the attacker's goals and the mechanism that stands against each. The details are in \cref{sec:design}.

\begin{table*}[t]
\centering\small
\caption{Attacker goals and what stands in the way (abridged from the project's security policy).}\label{tab:goals}
\begin{tabularx}{\textwidth}{@{}P{3.3cm}Y@{}}
\toprule
Goal & What stops it \\
\midrule
Run code on the host & the KVM boundary and Firecracker's small device model with its default seccomp filter (no jailer in v0); the daemon parses guest input as untrusted (bounded frames, strict JSON, path checks) \\
Reach the network directly & the VM has no network card; the only route out is the egress proxy on vsock \\
Reach internal services or cloud metadata & names resolved once on the host; loopback, private, link-local, metadata, multicast and the host's own addresses refused; the proxy dials the address it checked, so a name cannot rebind \\
Exfiltrate to an arbitrary host & the allowlist; any other host waits for a person on the host; TLS server name must equal the requested host; encrypted ClientHellos refused; a non-TLS tunnel only to an entry that names its port \\
Steal a secret & the value never enters the VM; the proxy swaps it into one bound header for exact bound hosts, blocks the placeholder or value elsewhere on traffic it reads, and redacts the value from responses \\
Take credentials from the checkout & the workspace builder never copies credential files (\code{.env*}, \code{*.pem}, \code{*.key}, \code{.netrc}, \code{.npmrc}, SSH keys, \code{.aws/}, \ldots) \\
Start a command without a record & commands started from the host run only through the guest agent's \code{exec}, decided by the signer before they start \\
Change the repository behind your back & changes leave only as a patch, applied by \code{tkbox apply} through a strict pipeline that decides every file before writing \\
Rewrite or silence the record & the record is in the signer's trust domain; Box fails closed; a must-keep record that cannot be recorded stops the box and is counted as \code{tkbox.gap} \\
Reach the daemon or signer sockets & both are Unix sockets on the host; the VM has no path to them \\
\bottomrule
\end{tabularx}
\end{table*}

\subsection{What Box guarantees}\label{sec:guarantees}

With a signer the agent cannot reach, and within the limits of \cref{sec:limits}:
\begin{enumerate}[leftmargin=*,label=\textbf{G\arabic*.}]
  \item Every command started in the box through \code{tkbox exec} (or the SDK, the GitHub Action or MCP) was decided by the signer before it started, and its outcome is in the signed record, or counted in the run's \code{tkbox.gap} record if it could not be recorded.
  \item Every outbound connection was decided before it was dialled: allowed by the allowlist, approved by a person on the host, or refused. Each is in the record, except Box's own refusals beyond 20 a second per box, which are counted in a summary record.
  \item A bound secret's value is sent only to its bound hosts, and only in its bound header. Each use is recorded by name and host, never by value.
  \item \code{tkbox apply} writes nothing to the checkout that the signer did not allow, file by file.
  \item The bundle is the signer's record of the run; \code{tracekit verify} detects any change to it.
\end{enumerate}

\begin{limitbox}[title={What Box does not guarantee}]
\begin{itemize}[leftmargin=*]
  \item \textbf{Processes inside the VM.} The processes a command starts itself are not decided or recorded one by one; they show up where they cross a door.
  \item \textbf{What an allowed host is sent.} Box decides who the agent may talk to, not what it says. A host that fetches on its caller's behalf (such as the Go module proxy) can relay workspace data further.
  \item \textbf{Leaks on uninspected connections.} On HTTPS hosts without a bound secret the proxy sees only the host, port and server name; only a placeholder could leave there, never a value.
  \item \textbf{Assurance.} A same-user dev signer gives \code{Assurance: dev}: the agent in the VM cannot reach its key, but another process of the same user could. Stronger assurance needs a signer under its own user and a witness on another host.
  \item \textbf{The host.} Root and the operator are trusted; v0 runs Firecracker without its jailer.
\end{itemize}
\end{limitbox}

\section{Design}\label{sec:design}

\subsection{A job's life}

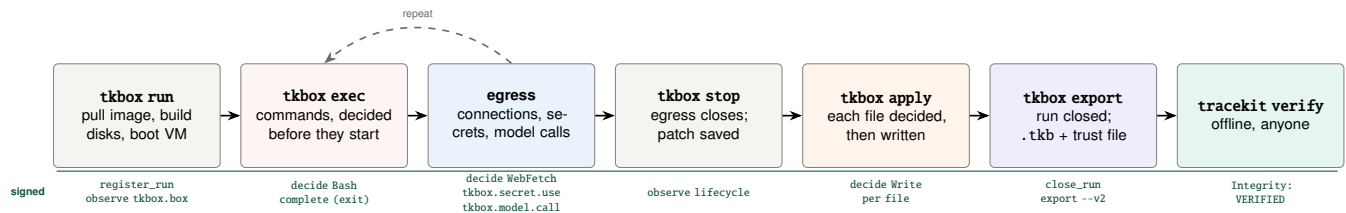
\begin{figure*}[t]
\centering
\resizebox{\textwidth}{!}{%
\begin{tikzpicture}[x=1cm,y=1cm,font=\sffamily\scriptsize,
  st/.style={draw=tkink!60,rounded corners=2pt,fill=white,align=center,text width=2.35cm,minimum height=1.55cm,inner sep=2pt},
  ar/.style={-{Stealth[length=1.9mm]},line width=0.7pt},
  rec/.style={font=\ttfamily\tiny,text=tkaqua!45!black,align=center,text width=2.35cm}]
\node[st,fill=tkfill] (r) at (0,0) {\textbf{\texttt{tkbox run}}\\pull image, build disks, boot VM};
\node[st,fill=tkredfill!50] (e) at (2.8,0) {\textbf{\texttt{tkbox exec}}\\commands, decided before they start};
\node[st,fill=tkbluefill] (n) at (5.6,0) {\textbf{egress}\\connections, secrets, model calls};
\node[st,fill=tkfill] (s) at (8.4,0) {\textbf{\texttt{tkbox stop}}\\egress closes; patch saved};
\node[st,fill=tkcream] (a) at (11.2,0) {\textbf{\texttt{tkbox apply}}\\each file decided, then written};
\node[st,fill=tkpurplefill] (x) at (14.0,0) {\textbf{\texttt{tkbox export}}\\run closed; \texttt{.tkb} + trust file};
\node[st,fill=tkaquafill] (v) at (16.8,0) {\textbf{\texttt{tracekit verify}}\\offline, anyone};
\foreach \p/\q in {r/e,e/n,n/s,s/a,a/x,x/v} \draw[ar] (\p) -- (\q);
\draw[ar,dashed,tkgrey] (n.north) to[out=140,in=40] node[above,font=\sffamily\tiny,text=tkgrey]{repeat} (e.north);
\node[rec] at (0,-1.15) {register\_run\\observe tkbox.box};
\node[rec] at (2.8,-1.15) {decide Bash\\complete (exit)};
\node[rec] at (5.6,-1.15) {decide WebFetch\\tkbox.secret.use\\tkbox.model.call};
\node[rec] at (8.4,-1.15) {observe lifecycle};
\node[rec] at (11.2,-1.15) {decide Write\\per file};
\node[rec] at (14.0,-1.15) {close\_run\\export --v2};
\node[rec] at (16.8,-1.15) {Integrity:\\VERIFIED};
\draw[tkaqua!60!black,line width=0.5pt] (-1.2,-0.83) -- (17.95,-0.83);
\node[font=\sffamily\tiny\bfseries,text=tkaqua!50!black,anchor=east] at (-1.25,-1.15) {signed};
\end{tikzpicture}}
\caption{The life of a job and what each step records in the box's Tracekit run (bottom row).}\label{fig:life}
\end{figure*}

\Cref{fig:life} shows a job from start to verification. \code{tkbox run} builds the box's disks, registers a Tracekit run and boots the VM; \code{tkbox exec} runs commands in it; every connection, secret use and model call passes the proxy; \code{tkbox stop} saves the workspace patch; \code{tkbox apply} writes the decided changes to the checkout; and \code{tkbox export} closes the run and writes the bundle and its trust file.

\subsection{The box}\label{sec:box}

\paragraph{No network card.} The VM is a Firecracker micro-VM~\cite{agache2020firecracker} with a pinned guest kernel (6.1.155) and VMM (v1.17.0), both downloaded once and checked by SHA-256. It has no network interface: its only device for talking to the host is vsock, which Firecracker exposes on the host as a Unix socket. The daemon serves three ports on it (\cref{tab:vsock}), and nothing else in the guest can reach the host or the network. Because there is no interface, there is no DNS and no UDP from the box at all: the guest cannot leak data in name lookups, because it cannot make any.

\begin{table*}[t]
\centering\small
\caption{The vsock ports: the only channels between guest and host.}\label{tab:vsock}
\begin{tabularx}{\textwidth}{@{}lP{2.3cm}Y@{}}
\toprule
Port & Direction & What travels \\
\midrule
1024 & host $\to$ guest & agent RPC: \code{exec}, \code{diff}, files, \code{shutdown}, in framed messages (\code{[type u8][len u32][payload]}, payload at most 1\,MiB) \\
1025 & guest $\to$ host & egress: raw HTTP proxy protocol from the guest's forwarder on \code{127.0.0.1:3128} to the box's proxy \\
1026 & guest $\to$ host & boot config: box id, environment, proxy address, the box's CA certificate, devices, limits, the host's clock; served once, then closed \\
\bottomrule
\end{tabularx}
\end{table*}

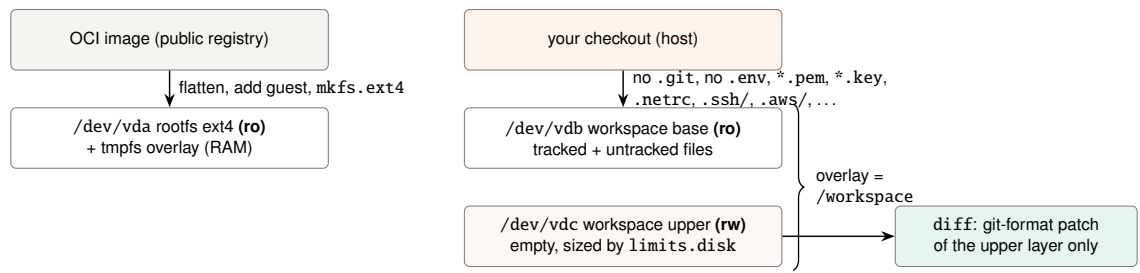
\begin{figure*}[t]
\centering
\begin{tikzpicture}[x=1cm,y=1cm,font=\sffamily\scriptsize,
  dk/.style={draw=tkink!60,rounded corners=2pt,align=center,minimum width=4.2cm,minimum height=0.8cm},
  ar/.style={-{Stealth[length=1.8mm]},line width=0.55pt}]
\node[dk,fill=tkfill] (img) at (0,2.4) {OCI image (public registry)};
\node[dk,fill=white] (vda) at (0,1.1) {\texttt{/dev/vda} rootfs ext4 \textbf{(ro)}\\+ tmpfs overlay (RAM)};
\node[dk,fill=tkcream] (co) at (6.0,2.4) {your checkout (host)};
\node[dk,fill=white] (vdb) at (6.0,1.1) {\texttt{/dev/vdb} workspace base \textbf{(ro)}\\tracked + untracked files};
\node[dk,fill=tkredfill!40] (vdc) at (6.0,-0.2) {\texttt{/dev/vdc} workspace upper \textbf{(rw)}\\empty, sized by \texttt{limits.disk}};
\node[dk,fill=tkaquafill,minimum width=3.2cm] (patch) at (11.2,-0.2) {\texttt{diff}: git-format patch\\of the upper layer only};
\draw[ar] (img) -- node[right]{flatten, add guest, \texttt{mkfs.ext4}} (vda);
\draw[ar] (co) -- node[right,align=left]{no \texttt{.git}, no \texttt{.env}, \texttt{*.pem}, \texttt{*.key},\\\texttt{.netrc}, \texttt{.ssh/}, \texttt{.aws/}, \ldots} (vdb);
\draw[ar] (vdc) -- (patch);
\draw[decorate,decoration={brace,amplitude=4pt,mirror},line width=0.5pt] (8.25,-0.65) -- node[right=5pt,align=left]{overlay =\\\texttt{/workspace}} (8.25,1.55);
\end{tikzpicture}
\caption{The box's disks. The patch is computed from the workspace's upper layer, so it contains exactly what the job changed.}\label{fig:disks}
\end{figure*}

\paragraph{Disks.} A box has three disks (\cref{fig:disks}). The root filesystem is built from a public OCI image by Box's own registry client: layers are flattened with whiteouts, the guest agent is added, and the result is written as an ext4 image and cached by manifest and guest-agent digest. It is mounted read-only with a tmpfs overlay, so writes outside the workspace live in RAM and vanish with the box. The workspace is an overlay of a read-only base disk, built from the checkout, and an empty writable upper disk. The base contains tracked and untracked-but-not-ignored files of a git work tree, never \code{.git}, and never credential files, ignored or not. Because the guest's \code{diff} walks only the upper layer, the patch is exact and cheap, and the image needs no git.

\paragraph{The guest is untrusted, including its agent.} \code{tkbox-guest} runs as PID~1 and serves exec, diff and the egress forwarder. The host never trusts what it says about itself: frames are bounded, JSON is parsed strictly, paths are checked, and only what passes a door is recorded as fact.

\subsection{Commands}\label{sec:commands}

A command started in the box from the host (\code{tkbox exec}, the Python SDK, the GitHub Action or the MCP server) runs only through the guest agent's \code{exec}, and only after the signer allows it. Box sends the signer a \code{Bash} decision whose arguments are the shell-quoted command line, the working directory, the exec's environment and its stdin when known up front (\cref{sec:record}). Three details make that record faithful:
\begin{itemize}[leftmargin=*]
  \item \textbf{The program that runs is the program the signer reads.} A first word that contains \code{=} or is a shell reserved word is quoted, so a shell parser sees the program that runs; a first word starting with \code{\%} (a job spec to bash) is refused.
  \item \textbf{The environment cannot change what runs unseen.} A per-exec environment appears in the decided command as a leading \code{env K=V \ldots}, and names that change what a program loads or runs (every \code{LD\_*}, \code{GIT\_*}, \code{PYTHON*}, \code{NODE\_*}, \code{BASH\_*}, \code{PATH}, \code{HOME}, \code{PS4}, build-tool wrappers and more, bar a few harmless ones) are refused per exec.
  \item \textbf{Stdin is decided when it can be.} Stdin known up front (at most 1\,MiB) is decided with the command as its SHA-256 and size, plus its text when it is UTF-8 of at most 4\,KiB; exactly that data becomes the process's stdin. Streamed stdin is known only afterwards, so its digest goes in the completion.
\end{itemize}
The processes a command starts itself, such as the shell commands an agent runs, are not decided one by one. Containing them is the VM's job; what they do is recorded where it crosses another door.

\subsection{Network}\label{sec:network}

\begin{figure}[t]
\centering
\begin{tikzpicture}[x=1cm,y=1cm,font=\sffamily\scriptsize,
  st/.style={draw=tkink!60,rounded corners=2pt,fill=white,align=center,text width=3.5cm,minimum height=0.62cm,inner sep=2pt},
  dec/.style={draw=tkorange,rounded corners=6pt,fill=tkcream,align=center,text width=3.5cm,minimum height=0.62cm,inner sep=2pt},
  bad/.style={draw=tkred,rounded corners=2pt,fill=tkredfill,align=left,text width=2.75cm,minimum height=0.55cm,inner sep=2pt,execute at begin node=\nohyph},
  good/.style={draw=tkaqua!70!black,rounded corners=2pt,fill=tkaquafill,align=center,text width=3.5cm,minimum height=0.62cm,inner sep=2pt},
  ar/.style={-{Stealth[length=1.6mm]},line width=0.5pt},
  lab/.style={font=\sffamily\tiny,inner sep=1pt}]
\node[st,fill=tkredfill!35] (g) at (0,0) {guest: \texttt{CONNECT host:port}\\or \texttt{GET http://host/}};
\node[dec] (pre) at (0,-1.15) {secret in the host name, \texttt{CONNECT} to port 80, or off the list with \texttt{inspect: all}?};
\node[dec] (lit) at (0,-2.45) {on the allowlist, or an IP literal?};
\node[st] (res) at (0,-3.55) {resolve once on the host; check every address};
\node[st,draw=tkaqua!70!black,line width=0.7pt] (sig) at (0,-4.75) {Tracekit signer decides \texttt{WebFetch}};
\node[dec] (onl) at (0,-5.95) {on the allowlist, or approved for this box?};
\node[st] (hold) at (0,-7.15) {held for a person on the host: \texttt{tkbox approve} (2\,min)};
\node[st] (res2) at (0,-8.35) {resolve now; check every address};
\node[st] (dial) at (0,-9.45) {dial the checked address};
\node[dec] (tls) at (0,-10.55) {TLS tunnel: the ClientHello names this host only, and no ECH?};
\node[good] (ok) at (0,-11.75) {bytes flow (read by the proxy if a secret is bound to the host)};
\node[bad] (r0) at (3.95,-1.15) {refused at once (a leak if a secret)};
\node[bad] (r1) at (3.95,-3.55) {refused before any decision: private, link-local, metadata, own address};
\node[bad] (r2) at (3.95,-4.75) {refused: the signer's rule};
\node[bad] (r3) at (3.95,-7.15) {rejected or no answer: refused, \texttt{tkbox.egress}};
\node[bad] (r4) at (3.95,-8.35) {refused: private, metadata, own address};
\node[bad] (r5) at (3.95,-10.55) {refused: TLS alert};
\draw[ar] (g) -- (pre);
\draw[ar] (pre) -- node[lab,right]{no} (lit);
\draw[ar] (pre) -- node[lab,above]{yes} (r0);
\draw[ar] (lit) -- node[lab,right]{yes} (res);
\draw[ar] (lit.west) -- ++(-0.45,0) |- node[lab,left,pos=0.25,align=right]{no: not\\resolved} (sig.west);
\draw[ar] (res) -- node[lab,right]{public} (sig);
\draw[ar] (res) -- (r1);
\draw[ar] (sig) -- node[lab,right]{allow or flag} (onl);
\draw[ar] (sig) -- node[lab,above]{deny} (r2);
\draw[ar] (onl) -- node[lab,right]{no} (hold);
\draw[ar] (onl.west) -- ++(-0.45,0) |- node[lab,left,pos=0.25]{yes} (dial.west);
\draw[ar] (hold) -- node[lab,right]{approved} (res2);
\draw[ar] (hold) -- (r3);
\draw[ar] (res2) -- node[lab,right]{public} (dial);
\draw[ar] (res2) -- (r4);
\draw[ar] (dial) -- (tls);
\draw[ar] (tls) -- node[lab,right]{yes, or plain HTTP} (ok);
\draw[ar] (tls) -- node[lab,above]{no} (r5);
\end{tikzpicture}
\caption{Deciding one outbound connection, in the order the proxy checks. A name on the allowlist (or an IP literal) is resolved once on the host and checked before the signer is asked; a name off the list is not resolved until it is allowed, because the lookup itself would carry whatever the guest encoded in the name to that domain's DNS servers. The proxy dials the address it checked, so a name cannot rebind between the check and the connect, and it checks the TLS ClientHello before it forwards any byte of the tunnel. An allowlisted name skips the person, never the signer.}\label{fig:egress}
\end{figure}
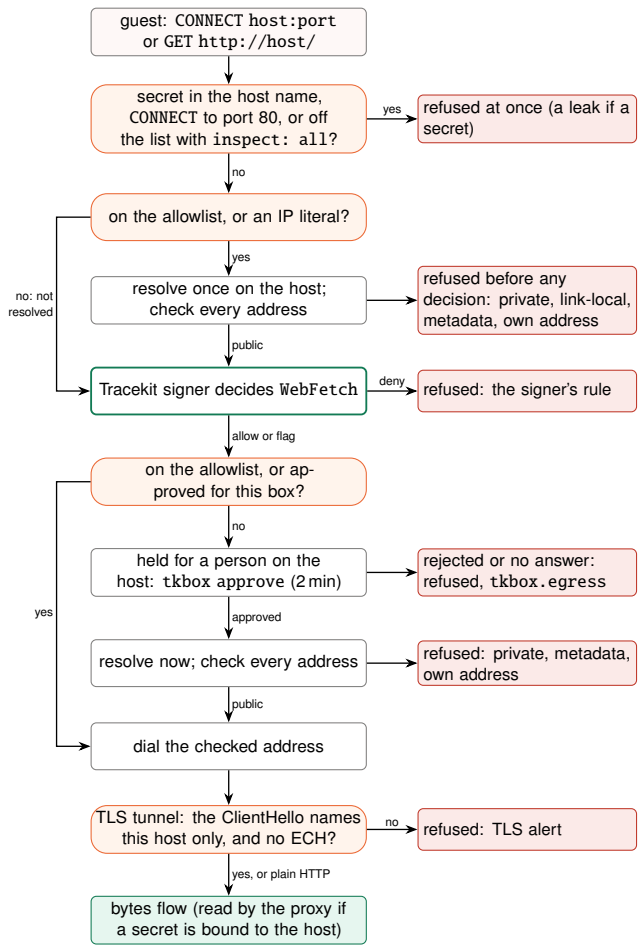

The guest's forwarder passes HTTP proxy traffic (\code{CONNECT} tunnels and absolute-form plain HTTP) over vsock to a per-box proxy on the host. \Cref{fig:egress} shows how one connection is decided.

\paragraph{Addresses.} The proxy resolves a name once, on the host, and refuses loopback, RFC~1918, CGNAT, unique-local, link-local, multicast, documentation and other special ranges, the cloud metadata addresses (169.254.169.254, fd00:ec2::254, 169.254.170.2, 100.100.100.200, 168.63.129.16 and others), and the host's own interface addresses, which it re-reads at most every 5\,s. IPv4-mapped and NAT64 addresses are unwrapped first. It then dials the address it checked. A name on the allowlist, or an IP literal, is resolved and checked before the signer is asked, so the metadata address is refused before any decision; a name off the allowlist is resolved only after the signer and a person have allowed it. On cloud VMs whose public IP is NAT'd and on no interface, the operator passes that address with \code{--egress-block}.

\paragraph{Allowlist and approvals.} An allow entry \code{host} permits HTTPS on 443 and plain HTTP on 80; \code{host:port} permits that port only; \code{*.host} permits subdomains. Presets name the hosts common jobs need (\cref{tab:presets}). Every connection is a \code{WebFetch} decision, on the list or off it. A host off the list is not refused outright: after the signer allows it, the connection waits up to two minutes for a person on the host to run \code{tkbox approve}, which nothing in the VM can do. At most 16 approvals may be pending and 30 opened per minute per box. With \code{inspect: all}, hosts off the list are refused at once.

\begin{table}[h]
\centering\footnotesize
\caption{Allow presets.}\label{tab:presets}
\begin{tabularx}{\linewidth}{@{}lY@{}}
\toprule
Preset & Hosts \\
\midrule
\code{pypi}, \code{npm} & \code{pypi.org}, \code{files.pythonhosted.org}; \code{registry.npmjs.org}, \code{registry.yarnpkg.com} \\
\code{github} & \code{github.com}, \code{api.github.com}, \code{codeload.github.com}, raw, release and LFS hosts \\
\code{anthropic}, \code{openai}, \code{claude-code} & \code{api.anthropic.com}; \code{api.openai.com}; \code{api.anthropic.com} and \code{statsig.anthropic.com} \\
\code{debian}, \code{apt-ubuntu} & Debian and Ubuntu package mirrors \\
\code{go-proxy} & \code{proxy.golang.org}, \code{sum.golang.org} (these fetch on the caller's behalf, so they can relay workspace data) \\
\bottomrule
\end{tabularx}
\end{table}

\paragraph{TLS.} Every HTTPS tunnel must start with a ClientHello whose single server name equals the \code{CONNECT} host, checked again after a HelloRetryRequest; encrypted ClientHellos are refused because Box must see the name. This stops fronting by SNI. A tunnel that is not TLS is allowed only to an entry that names its port, and is recorded as \code{tcp://host:port}, since the decision comes before the first byte. \code{CONNECT} to port 80 is refused, so plain HTTP always goes through the proxy in a form it reads.

\subsection{Secrets}\label{sec:secrets}

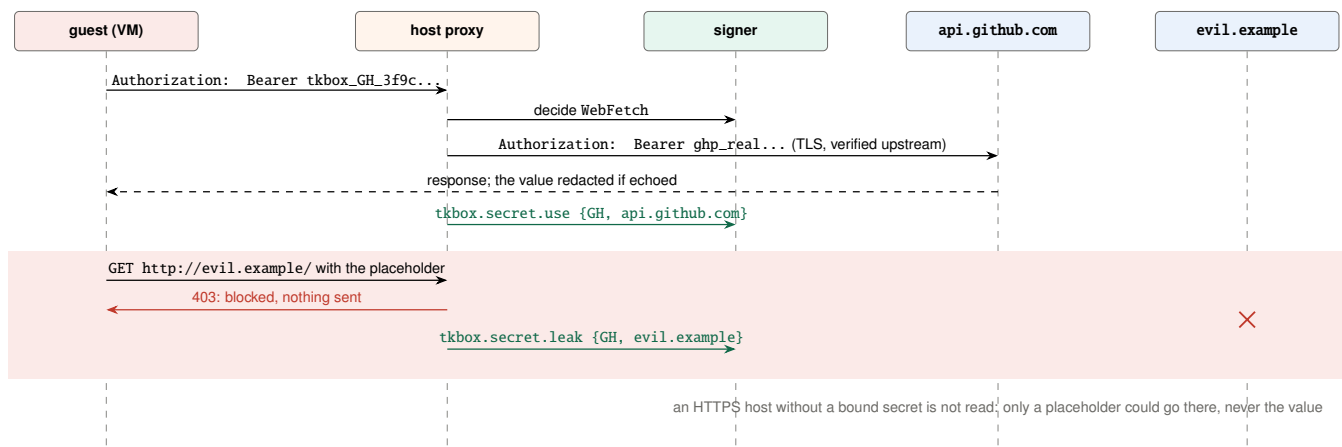
\begin{figure*}[t]
\centering
\resizebox{\textwidth}{!}{%
\begin{tikzpicture}[x=1cm,y=1cm,font=\sffamily\scriptsize,
  head/.style={draw=tkink!70,fill=white,rounded corners=2pt,minimum width=2.8cm,minimum height=0.6cm,align=center,font=\sffamily\bfseries\scriptsize},
  life/.style={draw=tkgrey!70,dashed,line width=0.5pt},
  msg/.style={-{Stealth[length=1.8mm]},line width=0.55pt},
  lab/.style={font=\sffamily\scriptsize,inner sep=1.5pt,align=center}]
\node[head,fill=tkredfill] (g) at (0,0) {guest (VM)};
\node[head,fill=tkcream] (p) at (5.2,0) {host proxy};
\node[head,fill=tkaquafill] (s) at (9.6,0) {signer};
\node[head,fill=tkbluefill] (b) at (13.6,0) {\texttt{api.github.com}};
\node[head,fill=tkbluefill] (e) at (17.4,0) {\texttt{evil.example}};
\foreach \n in {g,p,s,b,e} \draw[life] (\n.south) -- ++(0,-6.1);
\draw[msg] (0,-0.9) -- node[lab,above]{\texttt{Authorization: Bearer tkbox\_GH\_3f9c\ldots}} (5.2,-0.9);
\draw[msg] (5.2,-1.35) -- node[lab,above]{decide \texttt{WebFetch}} (9.6,-1.35);
\draw[msg] (5.2,-1.9) -- node[lab,above]{\texttt{Authorization: Bearer ghp\_real\ldots} (TLS, verified upstream)} (13.6,-1.9);
\draw[msg,dashed] (13.6,-2.45) -- node[lab,above]{response; the value redacted if echoed} (0,-2.45);
\draw[msg,tkaqua!60!black] (5.2,-2.95) -- node[lab,above]{\texttt{tkbox.secret.use \{GH, api.github.com\}}} (9.6,-2.95);
\fill[tkredfill] (-1.5,-3.35) rectangle (19,-5.3);
\draw[msg] (0,-3.8) -- node[lab,above]{\texttt{GET http://evil.example/} with the placeholder} (5.2,-3.8);
\draw[msg,tkred] (5.2,-4.25) -- node[lab,above]{403: blocked, nothing sent} (0,-4.25);
\draw[msg,tkaqua!60!black] (5.2,-4.85) -- node[lab,above]{\texttt{tkbox.secret.leak \{GH, evil.example\}}} (9.6,-4.85);
\draw[tkred,line width=0.8pt] (17.4,-4.4) node[cross out,draw,minimum size=6pt,inner sep=0pt]{};
\node[lab,text=tkgrey] at (13.6,-5.75) {an HTTPS host without a bound secret is not read: only a placeholder could go there, never the value};
\end{tikzpicture}}
\caption{A secret's two paths. To its bound host, in its bound header, the proxy swaps the placeholder for the real value and records the use. Anywhere else on traffic the proxy reads, the placeholder (or the value) is blocked and the attempt is recorded.}\label{fig:secret}
\end{figure*}

A secret is given on the host (\code{--secret GH=env:GH\_TOKEN@api.github.com}, or in \code{tkbox.yaml}), bound to exact host names (no wildcards) and to one header (\code{Authorization} by default). The VM sees only a placeholder, \code{tkbox\_GH\_} followed by 32 hex digits. \Cref{fig:secret} shows what happens to it.

On every connection the proxy reads (bound hosts, plain HTTP, and every allowed host with \code{inspect: all}) it blocks the placeholder anywhere but the bound header, and blocks the real value where it is not being swapped in, in its raw, percent-encoded, JSON-escaped, hex and base64 forms in any case. It redacts those forms from response headers, bodies and trailers, keeping their length so \code{Content-Length} still holds; requests to bound hosts ask for unencoded responses so they can be checked, and upgrades (WebSocket) and partial responses are refused on them. To read a bound host's TLS traffic the proxy uses a per-box ECDSA P-256 CA that is name-constrained to exactly those hosts, and verifies the upstream certificate against the host's roots. Before the box starts, Box also refuses any workspace file whose content, path or symlink target holds a secret's value, and any environment value or command line that does.

\subsection{Model calls}\label{sec:models}

A model API is egress to a model host named by a preset or allow entry (\code{--allow anthropic}), with the API key bound to it as a secret. The agent inside the box uses its normal SDK with the placeholder as its key. The connection gets its \code{WebFetch} decision like any other, and each request the proxy reads to a model host is also recorded as \code{tkbox.model.call} with host, path, status and size. A secret's host alone does not make a model host.

\subsection{Changes}\label{sec:changes}

\begin{figure*}[t]
\centering
\resizebox{0.98\linewidth}{!}{%
\begin{tikzpicture}[x=1cm,y=1cm,font=\sffamily\scriptsize,
  st/.style={draw=tkink!60,rounded corners=2pt,fill=white,align=center,text width=2.3cm,minimum height=1.6cm,inner sep=2pt},
  ar/.style={-{Stealth[length=1.8mm]},line width=0.6pt},
  bad/.style={font=\sffamily\scriptsize,text=tkred,align=center,text width=2.3cm}]
\node[st,fill=tkfill] (a) at (0,0) {\textbf{1 strict parse}\\every path checked; no copy or rename; one section per path};
\node[st] (b) at (2.75,0) {\textbf{2 git agrees}\\\texttt{git apply --numstat -z} reads the same paths, modes, binaries};
\node[st] (c) at (5.5,0) {\textbf{3 policy}\\no git or hooks dir; ignored files out; sensitive files need consent};
\node[st] (d) at (8.25,0) {\textbf{4 scratch build}\\\texttt{git apply} in a scratch copy, empty git dir and config};
\node[st,fill=tkcream] (e) at (11.0,0) {\textbf{5 decide}\\signer decides \texttt{Write} for every file: hash, mode, symlink};
\node[st] (f) at (13.75,0) {\textbf{6 re-check}\\results still match the decisions; checkout unchanged};
\node[st,fill=tkaquafill] (g) at (16.5,0) {\textbf{7 write}\\temp file, \texttt{O\_EXCL}, \texttt{O\_NOFOLLOW}, rename};
\foreach \p/\q in {a/b,b/c,c/d,d/e,e/f,f/g} \draw[ar] (\p) -- (\q);
\node[bad] at (0,-1.35) {unsafe path $\to$ refused};
\node[bad] at (2.75,-1.35) {git reads it differently $\to$ refused};
\node[bad] at (5.5,-1.35) {\texttt{.github/workflows/}\ldots{} without \texttt{--allow-}\allowbreak\texttt{sensitive} $\to$ refused};
\node[bad] at (8.25,-1.35) {does not apply $\to$ nothing written};
\node[bad] at (11.0,-1.35) {any file denied $\to$ nothing written};
\node[bad] at (13.75,-1.35) {mismatch $\to$ nothing written};
\node[bad,text=tkgrey] at (16.5,-1.35) {failure part way: written paths named, no rollback};
\end{tikzpicture}}
\caption{The \texttt{tkbox apply} pipeline. Nothing is written unless the signer allows every file, and what is written is what was decided.}\label{fig:apply}
\end{figure*}
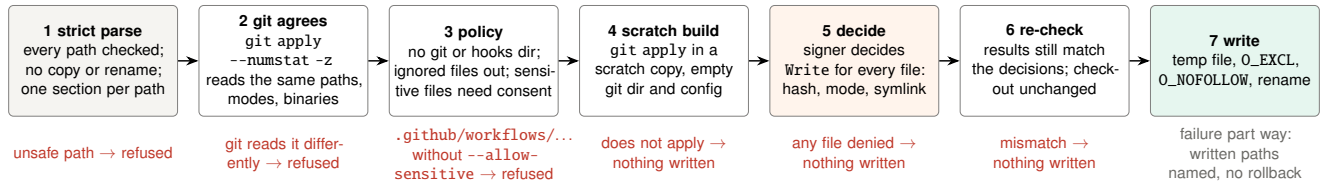

Changes leave the box only as a git-format patch of \code{/workspace}, computed in the guest from the overlay's upper layer and saved on the host when the box stops. \code{tkbox apply} writes it to the checkout, on a stopped box only, through the pipeline in \cref{fig:apply}. \code{git apply} never runs on the checkout.

Each step closes a specific gap between ``the patch the signer saw'' and ``the bytes on disk''. The strict parser and the cross-check against git's own reading (step 2, with an empty git directory and work tree) stop a patch that git would read differently from Box. The scratch build runs git with an empty \code{HOME}, no global or system config and no attributes file, so no filter, hook or LFS program of the user's runs on the agent's patch. Sensitive files, those that steer the next run or host automation, need \code{--allow-sensitive} (\cref{tab:sensitive}). The re-check in step 6 means a file changed on disk during the decisions is caught before anything is written, and Box writes each file itself, atomically, never through a symlinked parent.

\begin{table}[h]
\centering\footnotesize
\caption{Files a patch may change only with \texttt{--allow-sensitive}.}\label{tab:sensitive}
\begin{tabularx}{\linewidth}{@{}P{1.9cm}Y@{}}
\toprule
Kind & Files \\
\midrule
CI and automation & \code{.github/workflows/**}, \code{.github/actions/**}, \code{.gitlab-ci.yml}, \code{.circleci/**}, \code{.husky/**}, \code{lefthook.yml}, \code{.pre-commit-config.yaml} \\
Git behaviour & \code{.gitignore}, \code{.gitattributes}, \code{.gitmodules} \\
The next run & \code{tkbox.yaml}, \code{tkbox.yml}, \code{.envrc}, \code{.devcontainer/**}, \code{.vscode/tasks.json}, \code{.cargo/config.toml}, \code{.yarnrc.yml} \\
Credentials & \code{.npmrc}, \code{.netrc}, \code{*.pem}, \code{*.key} and the rest of the credential-file list \\
\bottomrule
\end{tabularx}
\end{table}

\subsection{Stop and containment}\label{sec:stop}

A box stops on its timeout (default one hour), on \code{tkbox stop}, when the signer holds or quarantines its run, or when Box fails closed. Egress closes the moment any stop begins: the proxy, its open connections and the egress listener close, and every later egress decision is refused, so no secret is swapped in while the box goes down. The daemon polls the signer's containment state for each running box (every 2\,s in its first minute, then every 15\,s, and at once after a decision that names an incident); a hold pauses the VM and a quarantine kills it.

\section{Recording the doors}\label{sec:record}

Box does not have its own log format, signer or verifier. It uses an unmodified Tracekit 1.4.2 signer through the signer's public RPC (newline-delimited JSON over its Unix socket, with the caller's identity taken from the socket's peer credentials), and Tracekit's bundle format and verifier~\cite{tracekit}. This section describes what Box sends to the signer, when, and what happens when the signer cannot take it.

\subsection{One run per box}

\begin{figure*}[t]
\centering
\begin{tikzpicture}[font=\sffamily\scriptsize,
  st/.style={draw=tkink!60,rounded corners=2pt,fill=white,minimum height=0.62cm,minimum width=1.55cm,align=center},
  rec/.style={draw=tkorange,fill=tkcream,rounded corners=2pt,minimum height=0.55cm,align=center,font=\ttfamily\scriptsize},
  ar/.style={-{Stealth[length=4pt]},thick,tkink!70}]
\node[st] (c) at (0,0) {created};
\node[st] (r) at (2.35,0) {running};
\node[st] (s) at (4.7,0) {stopped\\(patch saved)};
\node[st] (a) at (7.05,0) {applied};
\node[st,fill=tkfill] (e) at (9.4,0) {exported\\(run closed)};
\node[st,fill=tkredfill,draw=tkred!70] (q) at (4.7,-1.25) {failed /\\quarantined / rm};
\draw[ar] (c) -- (r);
\draw[ar] (r) -- node[above,font=\tiny]{stop} (s);
\draw[ar] (s) -- node[above,font=\tiny]{apply} (a);
\draw[ar] (a) -- node[above,font=\tiny]{export} (e);
\draw[ar] (s.north) to[out=40,in=140] node[above,font=\tiny]{export (no apply)} (e.north);
\draw[ar,tkred!70] (r.south) |- (q.west);
\draw[ar,tkred!70] (s) -- (q);
\node[rec] (k1) at (0,-2.5) {register\_run\\tkbox.box};
\node[rec] (k2) at (2.35,-2.5) {Bash, WebFetch\\observes};
\node[rec] (k3) at (7.05,-2.5) {Write\\per file};
\node[rec] (k4) at (9.4,-2.5) {close\_run\\trust + export};
\node[rec,draw=tkred,fill=tkredfill] (k5) at (4.7,-2.5) {close\_run\\(reason)};
\foreach \x/\y in {c/k1,r/k2,a/k3,e/k4,q/k5} \draw[densely dotted,tkgrey] (\x) -- (\y);
\end{tikzpicture}
\caption{A box's life and its Tracekit run. The run stays open after \code{tkbox stop} so that \code{tkbox apply} records its per-file \code{Write} decisions in the same run; export, removal, quarantine, a failure or daemon shutdown close it.}\label{fig:runlife}
\end{figure*}
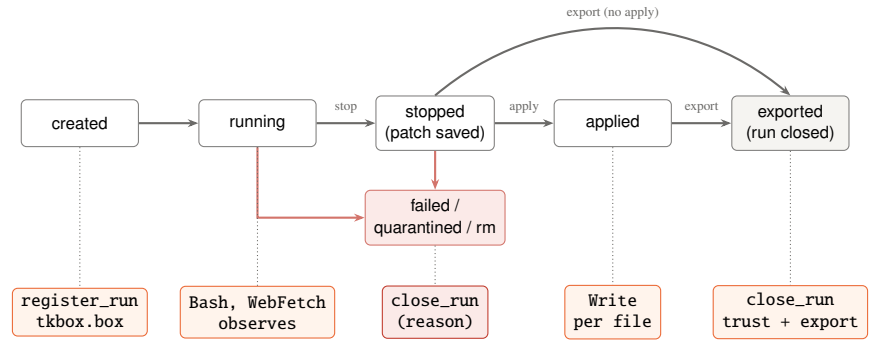

Each box is one Tracekit run (\cref{fig:runlife}). The daemon calls \code{register\_run} when the box is created and records the box's settings (image and its digest, allowlist, limits) as a trusted \code{tkbox.box} observation and the operator's task as a trusted \code{tkbox.task} observation, with every secret value and placeholder scrubbed. \code{tkbox stop} keeps the run open, so that the \code{Write} decisions of a later \code{tkbox apply} go into the same run as the commands and connections that produced the patch. \code{tkbox export} and \code{tkbox rm} close the run, as do daemon shutdown, quarantine and a failed box. Export and removal never close a run under a decision still in flight: they wait briefly for it, then refuse.

The run token is held only in the daemon's memory. After a daemon crash the run stays open in the signer, and that box can be neither applied nor exported: export refuses at once rather than produce a bundle the daemon cannot vouch for.

\subsection{From doors to records}

Tracekit records two kinds of entry: \emph{decisions}, where the signer evaluates a tool call against its policy before the call happens and later receives its completion, and \emph{observations}, which the signer signs but does not decide. \Cref{tab:doormap} gives the mapping for every door. Box maps its doors onto Tracekit's existing tool names (\code{Bash}, \code{WebFetch}, \code{Write}) so that the signer's standard policy packs apply to them without change; for example a \code{Bash} command that pipes a download into a shell, or a \code{Write} to a CI workflow file, meets the same rules it would meet from any other Tracekit adapter.

\begin{table*}[t]
\centering\small
\caption{What each door sends to the signer. Decisions are made before the action and completed after it; observations are signed but not decided. No record ever holds a secret's value.}\label{tab:doormap}
\begin{tabularx}{\textwidth}{@{}P{1.7cm}P{2.9cm}YP{3.6cm}@{}}
\toprule
Door & Kind & Arguments or fields & When \\
\midrule
Commands & decide \code{Bash} & \code{command} (the shell-quoted line, with per-exec env as a leading \code{env K=V}), \code{cwd}, \code{env} (placeholders shown as \code{tkbox\_NAME\_\ldots}), \code{stdin}; up-front stdin as \code{stdin\_sha256}, \code{stdin\_bytes} and, if UTF-8 and $\le$4\,KiB, \code{stdin\_text} & before the process starts; \code{complete} with the exit status (and streamed stdin's digest) \\
Network & decide \code{WebFetch} & \code{url}: \code{https://host:443}, \code{http://host:port} or \code{tcp://host:port} & per connection, before it is dialled \\
Approvals & observe \code{tkbox.approval} (trusted) & \code{\{approval\_id, target, decision, approver\}} & synchronously, before any connection uses an approval \\
Box's refusals & observe \code{tkbox.egress} & \code{\{host, port, decision, reason\}} for connections Box refused itself (metadata, private addresses, SNI mismatch, \ldots) & at the refusal; at most 20/s per box, then a summary \\
Secrets & observe \code{tkbox.secret.use} & \code{\{name, host\}} & on each swap into a bound header \\
 & observe \code{tkbox.secret.leak} & \code{\{name, host, reason\}} & on each blocked placeholder or value \\
Model calls & observe \code{tkbox.model.call} & \code{\{host, path, status, bytes\}} & per request to a model host the proxy reads \\
Changes & decide \code{Write} & \code{file\_path}, \code{sha256}, \code{mode}, \code{symlink}, \code{deleted}, \code{patch\_sha256}, and \code{sensitive: true} where it applies & for every file, before anything is written; completions carry \code{written\_paths} \\
Lifecycle & observe \code{tkbox.box} (trusted), \code{tkbox.task} (trusted) & box id, image and digest, allowlist, limits (scrubbed); the operator's task; agent choices are recorded separately as untrusted & synchronously, at each transition \\
Containment & observe \code{tkbox.contained} & \code{\{state, incident\_id\}} & when a hold or quarantine pauses or kills the box \\
Lost records & observe \code{tkbox.gap} & \code{\{kind, dropped, reason\}} & when a must-keep record could not be recorded \\
End & \code{close\_run} & the reason & export, rm, shutdown, quarantine, failure \\
\bottomrule
\end{tabularx}
\end{table*}

\paragraph{Commands.} The command line the signer sees is the one that runs. Box shell-quotes the argument vector itself, and quotes a first word that contains \code{=} or is a shell reserved word, so the signer's command parser sees the program that actually runs rather than an assignment or a keyword. Stdin that is known before the command starts (up to 1\,MiB) is decided together with the command, and exactly that data becomes the process's stdin.

\paragraph{Connections.} The \code{url} is fixed before the first byte of the connection, which is why a tunnel to a non-TLS port is recorded as \code{tcp://host:port}: Box cannot know what protocol will run over it, and does not claim to.

\paragraph{Approvals.} Tracekit 1.4.2 opens an approval only when policy returns \emph{ask}. A host off Box's allowlist is a Box rule, not a signer policy, so Box holds the connection itself and records the person's answer as \code{tkbox.approval}. An approval that allows a connection is recorded before the connection uses it; if the signer refuses that record, the connection is refused.

\paragraph{What the bundle holds.} In the exported bundle an observation's content appears as a salted HMAC with its size, as Tracekit records observations; the bundle proves that each observation was made, where it falls in the run, and that it was not changed. Each observation is labelled \emph{trusted} (Box's own statements: settings, task, approvals) or \emph{untrusted} (facts derived from what the guest sent, such as a connection attempt or a secret use), so a reviewer knows which entries rest on the agent's input.

\subsection{Failing closed}\label{sec:failclosed}

A record that is lost silently is worse than one that was never promised, because a reader of the bundle cannot tell the difference between ``did not happen'' and ``was not recorded''. Box's rules (\cref{tab:failclosed}) are designed so that every loss is either impossible or visible.

\begin{table}[t]
\centering\footnotesize
\caption{Fail-closed rules.}\label{tab:failclosed}
\begin{tabularx}{\linewidth}{@{}P{2.5cm}Y@{}}
\toprule
Situation & What Box does \\
\midrule
No signer at daemon start & no daemon, unless \code{--unrecorded}, which warns and records nothing \\
Signer stops answering & every later decision is refused; nothing runs, nothing connects \\
Must-keep observe cannot be queued (queue of 4096) or is refused & the box is stopped and the loss is recorded as \code{tkbox.gap} \\
Approval record refused & the connection is refused, the box stopped, the loss in \code{tkbox.gap}; if even that fails, the run is closed with the loss as its reason \\
A \code{complete} is refused or unanswered & counted in \code{tkbox.gap} with kind \code{complete}; the box keeps running \\
Box's own refusals above 20/s & a summary observation counts the rest \\
Daemon crash & the run stays open; the box cannot be applied or exported \\
Unrecorded box's patch on a recorded daemon & \code{tkbox apply} refuses it \\
\bottomrule
\end{tabularx}
\end{table}

``Must-keep'' observations are the ones a reviewer would need to judge the run: secret uses and leaks, model calls, Box's rejections and expiries of approvals, and containment. The only records Box rate-limits are its own refusals of connections (for example a loop hammering a refused host), and even those leave a count.

\subsection{Export and verification}

\begin{figure*}[t]
\centering
\includegraphics[width=0.72\textwidth]{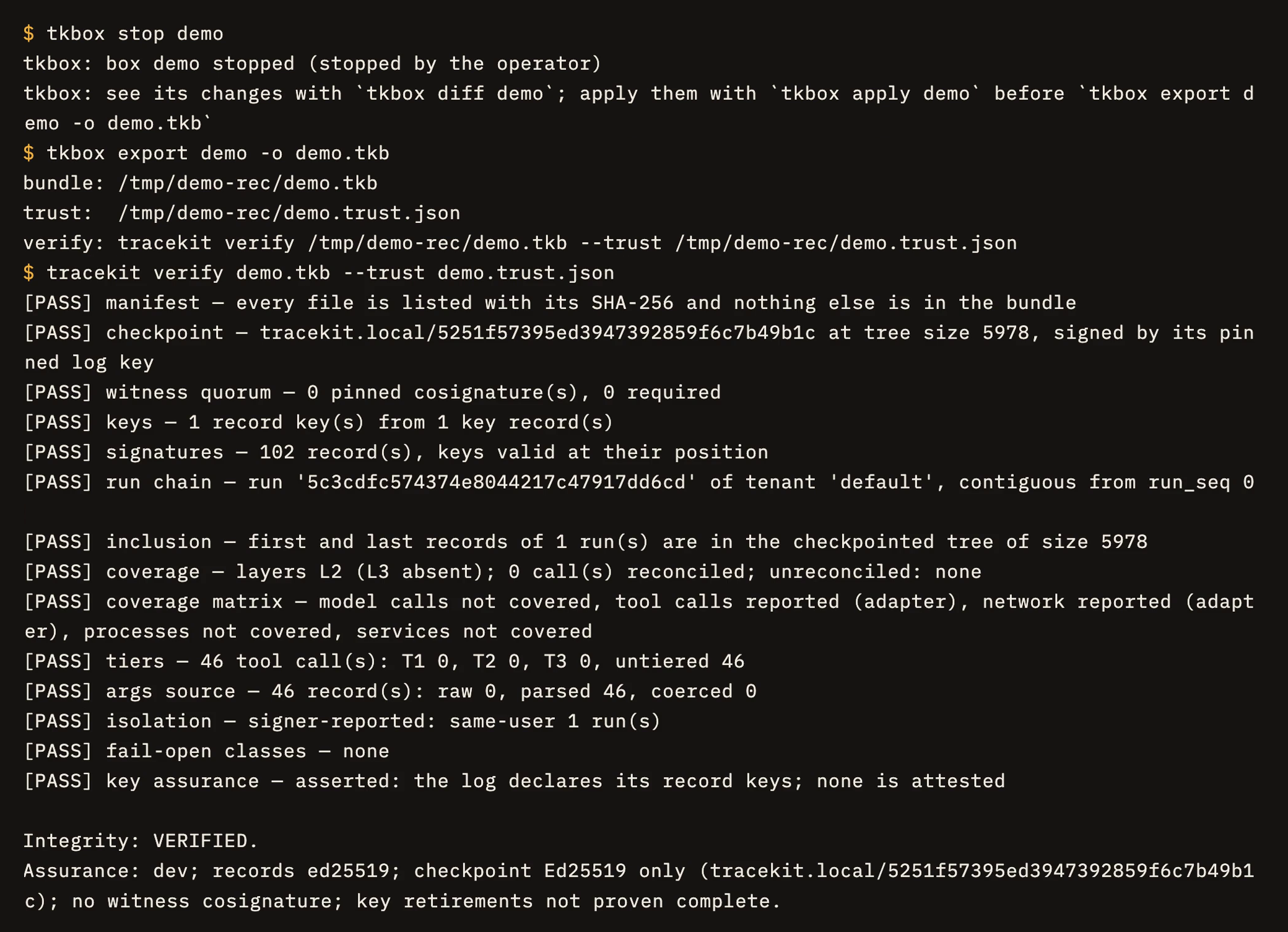}
\caption{Stopping, exporting and verifying the demo run, cut from the session's screen recording (lines over 104 columns wrapped). \code{tracekit verify} 1.4.2 checks the bundle offline: 102 signed records, a checkpoint at tree size 5,978, \code{Integrity: VERIFIED}, \code{Assurance: dev}. The coverage matrix is Tracekit's generic view: it does not know that in a box, the network and file writes have no other way out.}\label{fig:verify}
\end{figure*}

\code{tkbox export} closes the run, then runs \code{tracekit signer trust} and \code{tracekit export --v2}, which write the bundle \code{run.tkb} and its trust file \code{run.trust.json}. Anyone with both files can run \code{tracekit verify} offline: it checks the manifest, the signed checkpoint, every record signature, the run's hash chain and the inclusion of the run's first and last records in the checkpointed Merkle tree~\cite{rfc6962}. \Cref{fig:verify} shows the verifier's report for the demo run in the repository.

\subsection{Assurance}

What a verified bundle proves depends on who could reach the signer's key (\cref{tab:assurance}). The integrity check proves that no record was changed after it was signed, by anyone, in every row. What changes is who could have signed something false in the first place.

\begin{table}[t]
\centering\footnotesize
\caption{Assurance by deployment.}\label{tab:assurance}
\begin{tabularx}{\linewidth}{@{}P{1.95cm}P{2.0cm}Y@{}}
\toprule
Deployment & Verifier says & Who could forge a record \\
\midrule
Dev signer (\code{tracekit init --dev}), as the daemon's user & \code{Assurance: dev}, \code{same-user} & not the agent (it is in the VM), but any process of that user \\
Signer as its own user on the daemon's host & \code{separate-user} & root, or whoever controls the signer's user \\
Plus a witness on another host & above \code{dev} & a rewrite after a checkpoint is cosigned would be caught by the witness \\
GitHub Action & \code{Assurance: dev} & the job itself; the trust file shows only that the bundle was not changed after the job \\
\bottomrule
\end{tabularx}
\end{table}

All the bundles reported in this paper come from dev signers. They show that the doors were recorded and that the records were not changed afterwards; they do not show whose machine produced them.

\section{Implementation}\label{sec:impl}

Tracekit Box v0 is written in Go 1.26 and released under the MIT licence. It has two third-party Go dependencies (\code{golang.org/x/sys} and \code{gopkg.in/yaml.v3}); the egress proxy, the TLS ClientHello parser, the patch parser and the OCI image puller are written against the standard library. On the host it runs \code{mkfs.ext4} and \code{debugfs} to build disk images and \code{git} to read the checkout and cross-check patches. The repository holds about 25,600 lines of non-test Go and about 19,000 lines of Go tests (342 \code{Test} functions, of which 340 are unit tests and 2 are the end-to-end tests of \cref{sec:e2e}, and 15 \code{Fuzz} functions), plus a Python SDK (about 1,800 lines with its tests) and a GitHub Action.

\subsection{Structure}

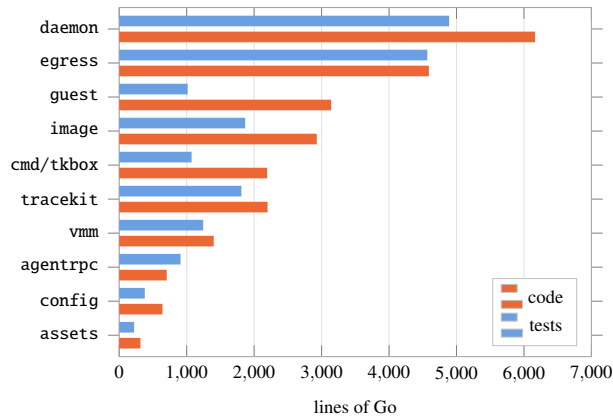
\begin{figure}[t]
\centering
\begin{tikzpicture}
\begin{axis}[xbar, width=0.92\linewidth, height=6.2cm, bar width=4pt,
  xlabel={lines of Go}, xmin=0, xmax=7000,
  symbolic y coords={assets,config,agentrpc,vmm,tracekit,cmdtkbox,image,guest,egress,daemon},
  ytick=data, yticklabels={assets,config,agentrpc,vmm,tracekit,cmd/tkbox,image,guest,egress,daemon},
  y tick label style={font=\scriptsize\ttfamily}, x tick label style={font=\scriptsize},
  xlabel style={font=\scriptsize}, legend style={font=\scriptsize,at={(0.97,0.03)},anchor=south east,draw=tkgrey!40},
  enlarge y limits=0.07, xmajorgrids, grid style={tkgrey!20}, axis line style={tkgrey!60}]
\addplot[fill=tkorange,draw=none] coordinates {(315,assets) (641,config) (704,agentrpc) (1401,vmm) (2199,tracekit) (2192,cmdtkbox) (2929,image) (3142,guest) (4590,egress) (6165,daemon)};
\addplot[fill=tkblue!70,draw=none] coordinates {(221,assets) (379,config) (909,agentrpc) (1245,vmm) (1811,tracekit) (1073,cmdtkbox) (1867,image) (1015,guest) (4568,egress) (4891,daemon)};
\legend{code, tests}
\end{axis}
\end{tikzpicture}
\caption{Code and test lines by package. The proxy (\code{egress}) is tested about line for line.}\label{fig:loc}
\end{figure}

\Cref{tab:packages} lists the packages and \cref{fig:loc} their sizes. Most of the code, and most of the tests, sit at the two doors an agent is most likely to attack: the daemon (which owns commands, approvals and \code{tkbox apply}) and the egress proxy (which owns connections, secrets and model calls). The egress package has about as many lines of tests as of code.

\begin{table*}[t]
\centering\small
\caption{Packages. Lines of non-test and test Go (\code{wc -l}) at commit \code{473723b}.}\label{tab:packages}
\begin{tabularx}{\textwidth}{@{}P{2.9cm}Yrr@{}}
\toprule
Package & Role & Code & Tests \\
\midrule
\code{internal/daemon} & box manager, HTTP API, exec, approvals, apply, events, containment & 6,165 & 4,891 \\
\code{internal/egress} & proxy, allowlist and presets, address checks, per-box CA, secret swap, leak blocking, redaction & 4,590 & 4,568 \\
\code{internal/guest} & PID~1 init, agent RPC server, exec, diff, egress forwarder, boot config & 3,142 & 1,015 \\
\code{internal/image} & OCI pull and cache, layer flattening, ext4 builds, the workspace image & 2,929 & 1,867 \\
\code{cmd/tkbox} & CLI and MCP server & 2,192 & 1,073 \\
\code{internal/tracekit} & signer RPC client, run lifecycle, decision gate, export & 2,199 & 1,811 \\
\code{internal/vmm} & Firecracker client, VM lifecycle, vsock & 1,401 & 1,245 \\
\code{internal/agentrpc} & host-side client of the guest agent & 704 & 909 \\
\code{internal/config} & \code{tkbox.yaml}, flags, limits & 641 & 379 \\
\code{cmd/tkbox-runs} & the Verified Runs table & 566 & 448 \\
\code{internal/proto}, \code{api} & frame format, guest operations, daemon API types & 709 & 221 \\
\code{internal/assets} & pinned Firecracker and kernel downloads & 315 & 221 \\
\code{e2e} & end-to-end suite on KVM & -- & 328 \\
\bottomrule
\end{tabularx}
\end{table*}

\paragraph{Testing without KVM.} Every boundary has an interface and a fake: \code{vmm.Launcher}, \code{agentrpc.Guest}, \code{image.Builder}, \code{egress.Decider} and \code{tracekit.Client}/\code{Run}. Packages therefore test without a VM, a registry or a signer, and the end-to-end suite (\cref{sec:e2e}) runs the real pieces together on a host with KVM.

\paragraph{Pinned assets.} On its first start the daemon downloads Firecracker v1.17.0 and guest kernel 6.1.155, checks each against a SHA-256 compiled into the binary, and caches them; a mismatch installs nothing. An operator can pass local files instead.

\paragraph{The guest.} \code{tkbox-guest} runs as PID~1. It mounts the root file system read-only with a tmpfs overlay, mounts the workspace overlay, reads its boot configuration once from vsock port 1026, starts the egress forwarder on \code{127.0.0.1:3128}, exports the proxy and CA variables to every command, and serves the agent RPC on port 1024. It is the only code Box puts inside the VM, and the daemon treats everything it says as untrusted: frames are bounded at 1\,MiB, JSON is parsed strictly, and paths are checked on the host.

\paragraph{State.} The daemon keeps images by digest and one directory per box (the Firecracker and vsock sockets, the console log, an events file capped at 256\,MiB, the two workspace disks, and the saved patch and its file summary). No file holds a secret value or the run token. The API is HTTP over a Unix socket that only the daemon's user and root can open.

\subsection{Interfaces}

Every interface goes through the same daemon API and the same doors; none has a path around a decision. \Cref{tab:cli} lists the CLI.

\begin{table}[t]
\centering\footnotesize
\caption{The \code{tkbox} command line.}\label{tab:cli}
\begin{tabularx}{\linewidth}{@{}P{3.6cm}Y@{}}
\toprule
Command & What it does \\
\midrule
\code{tkbox daemon} & run the daemon: boxes, egress proxies, the API socket \\
\code{tkbox run} & check the settings, then create and start a box \\
\code{tkbox exec <box> -- argv} & run a command in a box (decided first) \\
\code{tkbox ls} / \code{stop} / \code{rm} & list; stop and save the patch; delete \\
\code{tkbox diff <box>} & print the changes as a git-format patch \\
\code{tkbox apply <box>} & apply a stopped box's patch, file by file through the signer \\
\code{tkbox export <box> -o x.tkb} & close the run; write the bundle and trust file \\
\code{tkbox events <box>} & print the box's events \\
\code{tkbox image build <ref>} & pull an image and build its root file system \\
\code{tkbox approvals} / \code{approve} / \code{reject} & list and answer pending approvals \\
\code{tkbox mcp} & serve boxes as MCP tools on stdio \\
\bottomrule
\end{tabularx}
\end{table}

\paragraph{MCP server.} \code{tkbox mcp} lets an agent orchestrator on the host use boxes as tools (\code{tkbox\_run}, \code{\_exec}, \code{\_diff}, \code{\_stop}, \code{\_ls}). The operator's \code{--config} sets the image, allowlist, secrets, environment and limits, and a call can only \emph{narrow} them: a smaller allowlist drawn from the config's, an image the operator listed, a workspace under a root the operator named, resources at most the config's. A call that tries to add environment, set a task or name a box is refused with the field it would have widened. An MCP session reaches only the boxes it created, stops them when it ends, and has no tools for apply or export: a person does those. The narrowing holds only if the orchestrator cannot act as the daemon's user by another route, so the documentation tells operators to run it as a different user.

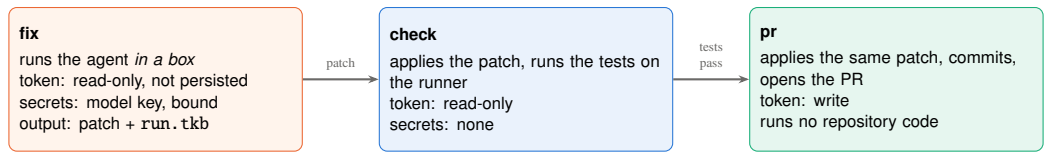
\begin{figure*}[t]
\centering
\begin{tikzpicture}[font=\sffamily\scriptsize,
  job/.style={draw=tkink!60,rounded corners=3pt,fill=white,text width=3.6cm,minimum height=1.9cm,align=left,inner sep=4pt},
  ar/.style={-{Stealth[length=4pt]},thick,tkink!70}]
\node[job,fill=tkcream,draw=tkorange] (f) at (0,0) {\textbf{fix}\\[2pt] runs the agent \emph{in a box}\\ token: read-only, not persisted\\ secrets: model key, bound\\ output: patch + \code{run.tkb}};
\node[job,fill=tkbluefill,draw=tkblue] (c) at (4.9,0) {\textbf{check}\\[2pt] applies the patch, runs the tests on the runner\\ token: read-only\\ secrets: none};
\node[job,fill=tkaquafill,draw=tkaqua] (p) at (9.8,0) {\textbf{pr}\\[2pt] applies the same patch, commits, opens the PR\\ token: write\\ runs no repository code};
\draw[ar] (f) -- node[above,font=\tiny,align=center]{patch} (c);
\draw[ar] (c) -- node[above,font=\tiny,align=center]{tests\\pass} (p);
\end{tikzpicture}
\caption{The three-job shape of the example workflow: only the box runs agent code while a model key is present, the job that runs the agent's code holds no secrets, and the job that holds the write token runs no repository code.}\label{fig:action}
\end{figure*}

\paragraph{GitHub Action.} The action runs Claude Code, Codex or a command in a box on a GitHub-hosted runner, applies the changes when the agent exits 0 and the signer allows every file, and uploads the bundle, trust file and patch. It never passes \code{--allow-sensitive}. Because code the agent wrote runs wherever it is run, the example workflow splits the job three ways (\cref{fig:action}).

\paragraph{Python SDK.} A standard-library client (Python 3.9 and later) for the daemon's API: \code{create\_box}, \code{exec}, \code{diff}, \code{apply}, \code{export}, \code{events}, \code{approvals}, \code{approve}, \code{reject}. A \code{Secret} carries its value and bound hosts to the daemon, which keeps the value out of the VM; its \code{repr} never prints the value. A daemon the client cannot reach raises \code{DaemonUnreachable}; a signer the daemon cannot reach raises \code{Unavailable}, because the daemon fails closed.

\paragraph{Verified Runs.} The repository keeps a folder of real agent runs, each a bundle, its trust file and the contributor's \code{run.yaml}. \code{make runs} verifies every bundle with \code{tracekit verify} and rewrites the table from the verified records (commands, connections, held, refused); a run that does not verify is not listed, and no number in the table is typed by hand.

\section{Evaluation}\label{sec:eval}

We evaluate whether the v0 implementation does what \cref{sec:guarantees} says, on real hardware, and how it has been tested. We ask four questions:
\begin{enumerate}[leftmargin=*,label=\textbf{Q\arabic*.}]
  \item Does each door behave as designed on a real KVM host, end to end? (\cref{sec:e2e})
  \item What does a signed record of a real box look like, and does it verify? (\cref{sec:demo})
  \item How well are the parsers at the trust boundary tested against hostile input? (\cref{sec:fuzz})
  \item What did review find, and what is still open? (\cref{sec:review})
\end{enumerate}

\begin{limitbox}[title={What this evaluation does not measure}]
\begin{itemize}[leftmargin=*]
  \item \textbf{Performance.} We report the end-to-end suite's wall-clock times, which include image pull and boot, but we have not benchmarked boot latency, exec overhead, proxy throughput or the cost of a signer decision per connection.
  \item \textbf{Adversarial agents.} No study has yet put a capable agent in a box with instructions to get out. The end-to-end checks are scripted attacks on each door, not an agent searching for one.
  \item \textbf{Deployments above dev assurance.} Every bundle here comes from a dev signer running as the daemon's user (\cref{tab:assurance}).
  \item \textbf{Scale.} One host, one daemon, a handful of boxes.
\end{itemize}
\end{limitbox}

\subsection{End to end on KVM}\label{sec:e2e}

\paragraph{Setup.} The \code{e2e} package runs the real daemon, real Firecracker, the real guest, a real Tracekit 1.4.2 dev signer and real registries, on a GitHub-hosted Ubuntu runner with \code{/dev/kvm}. It is the \code{e2e} job of the repository's \code{ci} workflow and runs on every push. We report CI run~\#16 (commit \code{473723b}, 26 October 2026): the job took 1\,min\,6\,s, of which \code{make e2e} took 33\,s.

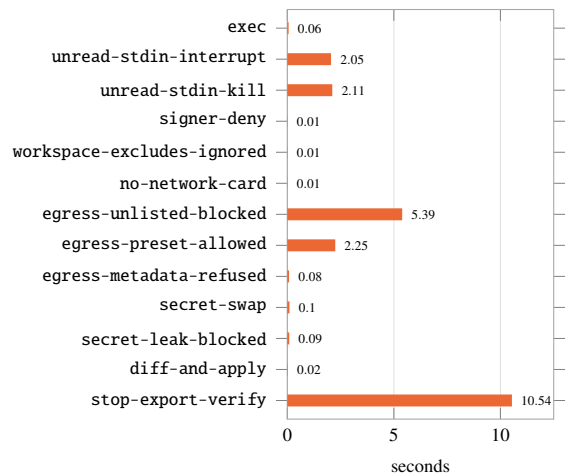
\begin{figure}[t]
\centering
\begin{tikzpicture}
\begin{axis}[xbar, width=0.6\linewidth, height=7.0cm, bar width=4.5pt,
  xlabel={seconds}, xmin=0, xmax=12.5,
  symbolic y coords={exec,unread-stdin-interrupt,unread-stdin-kill,signer-deny,workspace-excludes-ignored,no-network-card,egress-unlisted-blocked,egress-preset-allowed,egress-metadata-refused,secret-swap,secret-leak-blocked,diff-and-apply,stop-export-verify},
  ytick=data, y dir=reverse,
  y tick label style={font=\scriptsize\ttfamily}, x tick label style={font=\scriptsize}, xlabel style={font=\scriptsize},
  nodes near coords={\pgfmathprintnumber[fixed,precision=2]{\pgfplotspointmeta}},
  nodes near coords style={font=\tiny,anchor=west},
  enlarge y limits=0.05, xmajorgrids, grid style={tkgrey!20}, axis line style={tkgrey!60}]
\addplot[fill=tkorange,draw=none] coordinates {(0.06,exec) (2.05,unread-stdin-interrupt) (2.11,unread-stdin-kill) (0.01,signer-deny) (0.01,workspace-excludes-ignored) (0.01,no-network-card) (5.39,egress-unlisted-blocked) (2.25,egress-preset-allowed) (0.08,egress-metadata-refused) (0.1,secret-swap) (0.09,secret-leak-blocked) (0.02,diff-and-apply) (10.54,stop-export-verify)};
\end{axis}
\end{tikzpicture}
\caption{Duration of each end-to-end check in CI run \#16. The long bars are deliberate waits (a client giving up on a held host; commands that must be seen running before they are interrupted), a real package install, and a full stop--apply--export--verify.}\label{fig:e2etimes}
\end{figure}

\paragraph{Results.} \Cref{tab:e2e} lists the 13 checks of \code{TestBox}; all passed, in 28.47\,s for the test function. \code{TestFailClosed}, which checks that the daemon refuses to start without a signer, and does so before downloading anything, also passed. \Cref{fig:e2etimes} shows where the time goes: three checks wait on purpose (a client giving up on a held host, a command that must be seen to start before it is interrupted or its client killed), one installs a package from PyPI, and one stops a box, applies its patch, exports it and verifies the bundle. The checks of the doors themselves take tens of milliseconds.

\begin{table*}[t]
\centering\small
\caption{The end-to-end checks of \code{TestBox} in CI run \#16. All passed. Times are as \code{go test -v} printed them.}\label{tab:e2e}
\begin{tabularx}{\textwidth}{@{}P{4.1cm}P{1.7cm}Yr@{}}
\toprule
Check & Door & What it does and what it requires & Time (s) \\
\midrule
\code{exec} & commands & runs a command through the guest agent, decided by the signer & 0.06 \\
\code{unread-stdin-interrupt} & commands & a command with 2\,MB of stdin it never reads still stops when the client sends Ctrl-C & 2.05 \\
\code{unread-stdin-kill} & commands & \ldots{} and when the client is killed outright & 2.11 \\
\code{signer-deny} & commands & a command that would kill the signer (\code{pkill tracekitd}) is denied by the signer's policy before it starts: exit 126, rule \code{TK-D007} & 0.01 \\
\code{workspace-excludes-}\allowbreak\code{ignored} & workspace & the checkout's \code{.env} and \code{.git} are not in \code{/workspace}; its tracked files are & 0.01 \\
\code{no-network-card} & network & \code{/proc/net/dev} in the guest lists no interface but \code{lo} & 0.01 \\
\code{egress-unlisted-blocked} & network & a fetch of \code{example.com}, off the allowlist, is held for \code{tkbox approve}; unanswered, it fails & 5.39 \\
\code{egress-preset-allowed} & network & \code{pip install requests} succeeds through the \code{pypi} preset & 2.25 \\
\code{egress-metadata-refused} & network & a fetch of \code{169.254.169.254} fails: the proxy answers 403, ``a cloud metadata service address'' & 0.08 \\
\code{secret-swap} & secrets & a test server on the host, the secret's bound host, receives the real value in \code{Authorization} and answers \code{auth-ok}; inside the VM \code{\$TK} is only \code{tkbox\_TK\_\ldots} & 0.10 \\
\code{secret-leak-blocked} & secrets & the same placeholder in an \code{X-Smuggle} header is blocked with 403 & 0.09 \\
\code{diff-and-apply} & changes & a new file and an edit made in the box both appear in \code{tkbox diff} & 0.02 \\
\code{stop-export-verify} & changes, record & \code{tkbox stop}; \code{tkbox apply} writes the new file to the checkout; \code{tkbox export}; \code{tracekit verify} prints \code{VERIFIED} (64 records, tree size 65, 27 tool calls, \code{Assurance: dev}, \code{isolation: same-user}) & 10.54 \\
\midrule
\multicolumn{3}{@{}l}{\code{TestBox} total (including image pull and boot)} & 28.47 \\
\bottomrule
\end{tabularx}
\end{table*}

\paragraph{Reading the results against the guarantees.} \code{exec}, \code{signer-deny} and the stdin checks exercise G1; the three \code{egress-*} checks and \code{no-network-card} exercise G2; \code{secret-swap} and \code{secret-leak-blocked} exercise G3; the apply step of \code{stop-export-verify} exercises G4, and its export and verify exercise G5; \code{TestFailClosed} exercises the first fail-closed rule of \cref{tab:failclosed}. Each check exercises one representative case of its guarantee, not every case: the many forms of a private address, of an encoded secret or of a hostile patch are covered by unit and fuzz tests (\cref{sec:fuzz}), not end to end.

\paragraph{CI history.} The \code{ci} workflow (unit, end-to-end and action jobs) had run 16 times at the time of writing; four runs (\#1, \#2, \#5 and \#15) failed early, while the workflow and the end-to-end setup were being brought up, and the rest passed. A separate \code{fuzz} workflow runs every fuzz target nightly.

\subsection{A real box and its record}\label{sec:demo}

\begin{figure*}[p]
\centering
\begin{subfigure}{0.9\textwidth}\centering
\includegraphics[width=\linewidth]{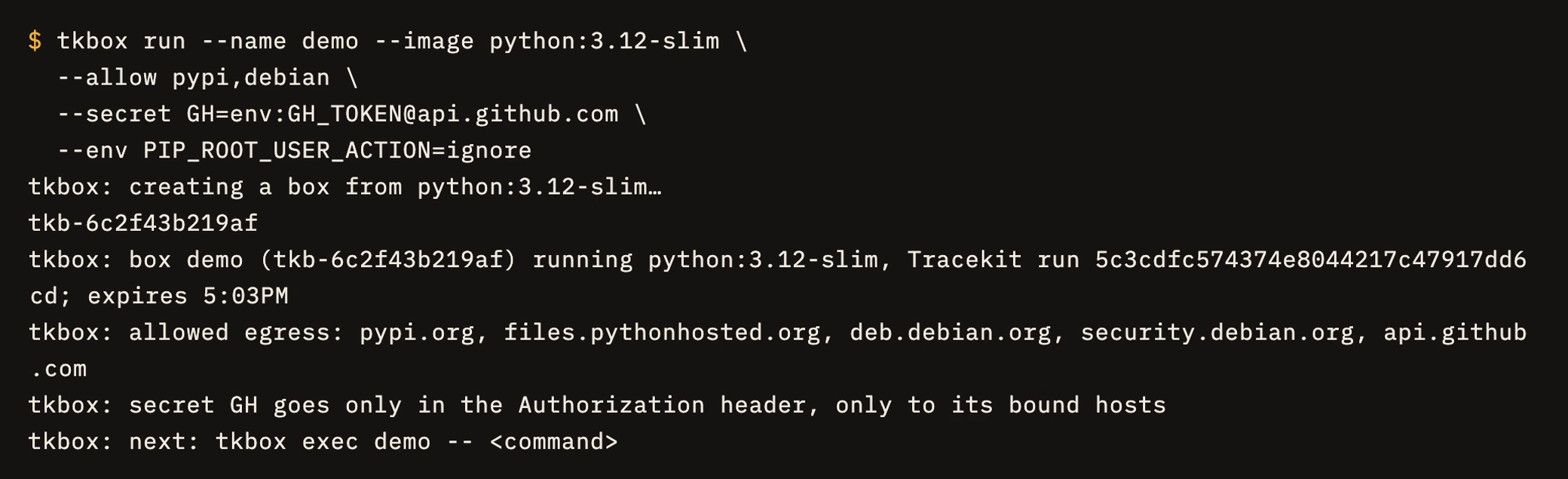}
\caption{\code{tkbox run} creates the box and registers its Tracekit run (\code{5c3cdfc5\ldots}, the run in the bundle). Box prints the hosts the allowlist expands to and the one header the secret may go in.}
\end{subfigure}\\[8pt]
\begin{subfigure}{0.9\textwidth}\centering
\includegraphics[width=\linewidth]{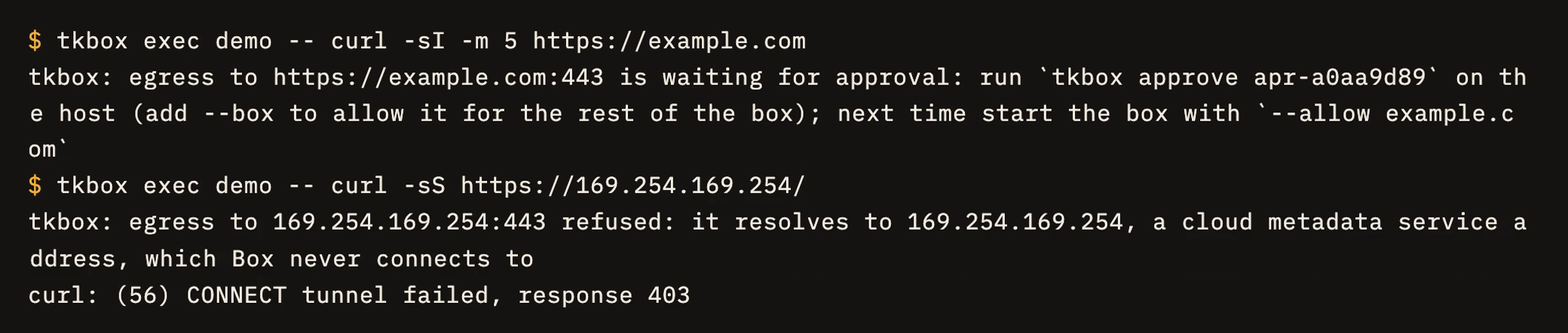}
\caption{A host off the allowlist is held for a person on the host (\code{tkbox approve}); \code{curl} gave up after 5\,s and no one approved it. The cloud metadata address is refused before any decision, with 403.}
\end{subfigure}\\[8pt]
\begin{subfigure}{0.9\textwidth}\centering
\includegraphics[width=\linewidth]{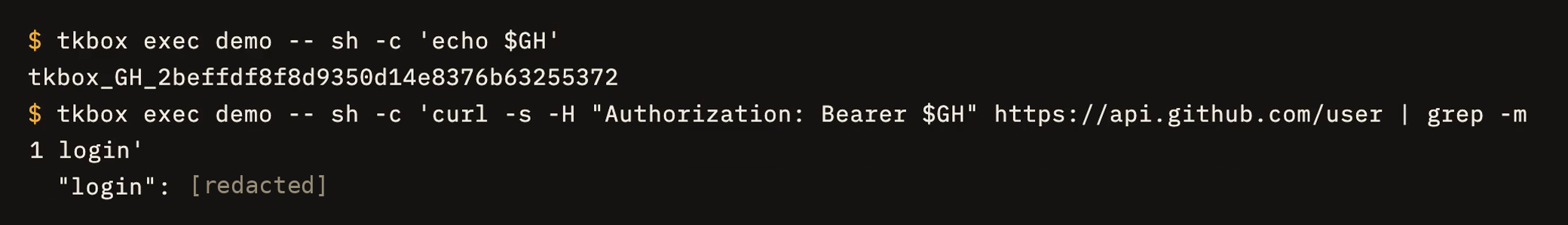}
\caption{Inside the box \code{\$GH} is only a placeholder. Sent in the \code{Authorization} header to \code{api.github.com}, its bound host, it works: the proxy swapped the real token in. (The account name in GitHub's answer is redacted here.)}
\end{subfigure}\\[8pt]
\begin{subfigure}{0.9\textwidth}\centering
\includegraphics[width=\linewidth]{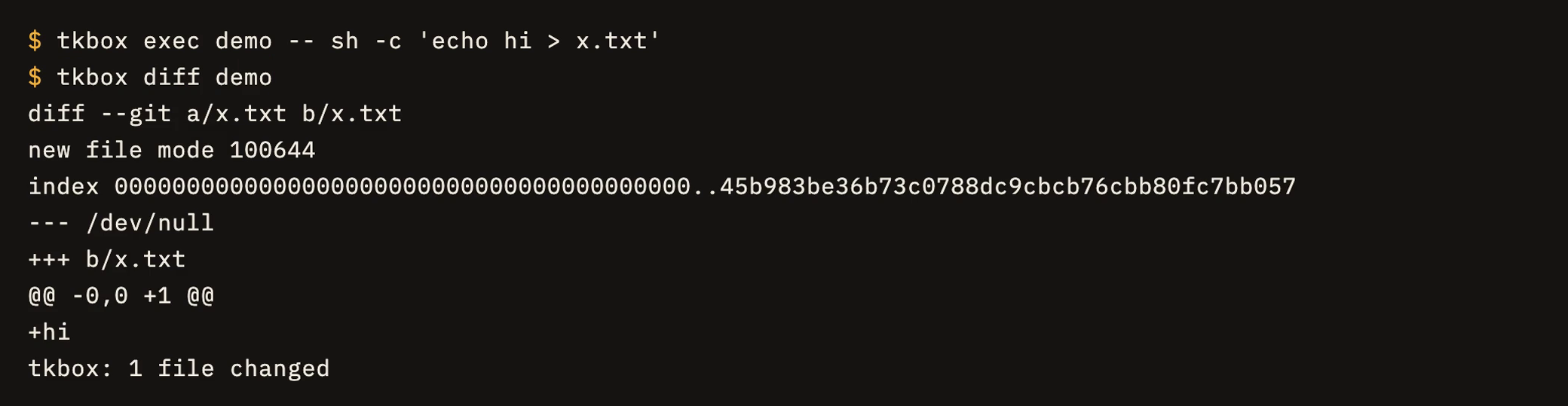}
\caption{A file written in the box comes back only as a git-format patch.}
\end{subfigure}
\caption{The demo run, as recorded. Each panel is cut from the screen recording of the session that produced the bundle in \cref{tab:demo}; lines longer than 104 columns are wrapped, and nothing else is changed except the redaction in (c). \Cref{fig:verify} shows the same session's stop, export and verify.}\label{fig:demo}
\end{figure*}

The repository's Verified Runs folder holds the run behind its README demo; \cref{fig:demo,fig:verify} show the terminal session that produced it, cut from its screen recording. An operator ran \code{python:3.12-slim} with \code{--allow pypi,debian} and one secret, \code{GH}, bound to \code{api.github.com}, and then, by hand with \code{tkbox exec}, installed \code{curl} (off camera, through the \code{debian} preset) and \code{requests} and tried each way out: a host off the allowlist (\code{example.com}), the cloud metadata address, and the GitHub token sent to its bound host.

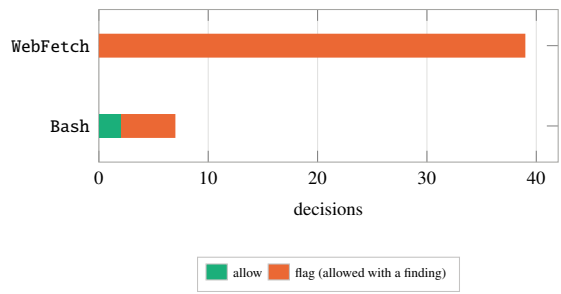
\begin{figure}[t]
\centering
\begin{tikzpicture}
\begin{axis}[xbar stacked, width=0.9\linewidth, height=3.6cm, bar width=9pt,
  xlabel={decisions}, xmin=0, xmax=42,
  symbolic y coords={Bash,WebFetch}, ytick=data,
  y tick label style={font=\scriptsize\ttfamily}, x tick label style={font=\scriptsize}, xlabel style={font=\scriptsize},
  legend style={font=\tiny,at={(0.5,-0.62)},anchor=north,draw=tkgrey!40,legend columns=2},
  enlarge y limits=0.45, xmajorgrids, grid style={tkgrey!20}, axis line style={tkgrey!60}]
\addplot[fill=tkaqua,draw=none] coordinates {(2,Bash) (0,WebFetch)};
\addplot[fill=tkorange,draw=none] coordinates {(5,Bash) (39,WebFetch)};
\legend{allow, flag (allowed with a finding)}
\end{axis}
\end{tikzpicture}
\caption{The demo run's 46 signer decisions. Every \code{WebFetch} was flagged by rules \code{TK-F002} and \code{TK-F012} of the signer's default packs; a flagged call is allowed and carries the finding in its record. The 39 include \code{example.com}: the signer decided it first, then Box held it for a person, and when no one approved it Box refused it and recorded \code{tkbox.egress}. The metadata address is not among them: Box refused it before any decision.}\label{fig:demochart}
\end{figure}

\Cref{tab:demo} breaks down the 102 records of the bundle; \cref{fig:demochart} shows the decisions. The bundle verifies with Tracekit 1.4.2 (\cref{fig:verify}).

\begin{table}[t]
\centering\footnotesize
\caption{The demo run's signed record (\code{runs/by-hand/readme-demo}).}\label{tab:demo}
\begin{tabularx}{\linewidth}{@{}Yr@{}}
\toprule
Record & Count \\
\midrule
\code{policy.decision} (Bash and WebFetch) & 46 \\
\code{tool.result} (43 ok, 3 not ok) & 46 \\
\code{input.observed}: \code{tkbox.box} & 3 \\
\quad \code{tkbox.egress} (Box's own refusals) & 2 \\
\quad \code{tkbox.secret.use} & 1 \\
\quad \code{tkbox.approval} & 1 \\
run lifecycle (\code{run.registered}, \code{run.closing}, \code{run.final}) & 3 \\
\midrule
Total records & 102 \\
\midrule
Verified Runs table: commands / connections / held / refused & 7 / 39 / 1 / 2 \\
Checkpoint tree size & 5,978 \\
Integrity & \code{VERIFIED} \\
Assurance; isolation & dev; same-user \\
\bottomrule
\end{tabularx}
\end{table}

Three details of this record are worth noting. First, the counts in the Verified Runs table are computed from the verified records by \code{make runs}, not typed: 7 commands, 39 connections, 1 held, 2 refused. Second, the cloud metadata address never reached the signer: Box refused it at the door and recorded \code{tkbox.egress}, so it is counted under refused but not under connections. Third, the held host (\code{example.com}) appears both as held and refused, because no one approved it and the hold ended in a refusal that Box recorded. Those two refusals are the two \code{tkbox.egress} observations in \cref{tab:demo}; \code{example.com} also has its \code{WebFetch} decision, because the signer is asked before a person is. The record shows exactly one secret use, by name and host, and the token's value appears nowhere in the bundle.

\paragraph{Coverage, as Tracekit sees it.} The verifier's coverage matrix (\cref{fig:verify}) reports network and tool calls as ``reported (adapter)'' and processes as ``not covered''. That is Tracekit's generic, conservative view: it does not know that inside a box the network has no other route. The record is complete for the doors because of the box, not because of anything the bundle can prove on its own; a reader who trusts the bundle must also trust that it came from a box, which the \code{tkbox.box} observation states but cannot prove at dev assurance.

\subsection{Fuzzing the trust boundary}\label{sec:fuzz}

Every parser that reads bytes from the guest, or from an upstream host on the guest's behalf, is a place where the guest can attack the host. \Cref{tab:fuzz} lists the 15 fuzz targets. Each is stated as a property, not just ``does not crash''; for example \code{FuzzClientHello} requires that a ClientHello Box admits names the \code{CONNECT} host, and only it, under three different readings of the same bytes (Go's \code{crypto/tls}, Box's own parser, and a reading that takes every \code{server\_name} entry of any type).

\begin{table*}[t]
\centering\small
\caption{The 15 fuzz targets and the property each one checks.}\label{tab:fuzz}
\begin{tabularx}{\textwidth}{@{}P{3.4cm}P{2.0cm}Y@{}}
\toprule
Target & Package & Property \\
\midrule
\code{FuzzAddrForms} & egress & any textual form of a blocked address is refused wherever the proxy parses a host; metadata addresses are refused even under \code{AllowPrivate}; block prefixes hold in every form \\
\code{FuzzHostText} & egress & host text accepted as an IP literal is checked as that literal; text accepted as a name is not one \code{inet\_aton} reads as IPv4 \\
\code{FuzzClientHello} & egress & an admitted ClientHello names the \code{CONNECT} host and only it, in every reading; exactly one ClientHello is forwarded \\
\code{FuzzForward} & egress & the upstream gets the path and query Box checked and recorded; the guest never sees a secret value a bound host echoed \\
\code{FuzzRedactReader} & egress & a body carrying secret values in any claimed form, split at any read boundary, comes out the same length and with none of them; gzip and deflate too \\
\code{FuzzReadFrame} & proto & any byte stream decodes to bounded frames, allocates at most one bounded payload per frame, and re-encodes exactly \\
\code{FuzzChunkWriter} & proto & any write splits into bounded frames that concatenate back to the input \\
\code{FuzzClientReplies} & agentrpc & whatever a hostile guest answers, the host client returns promptly, stays within its caps, and accepts a diff only with clean, relative, \code{.git}-free paths of the declared size \\
\code{FuzzParsePatch} & daemon & no panic; every path apply would decide and write is relative, clean and inside the root \\
\code{FuzzParsePatchVsGit} & daemon & when the parser accepts a patch, \code{git apply --numstat} names no path the parser did not return \\
\code{FuzzShellQuote} & tracekit & the signed command line parses back to exactly the argument vector, and its first word is the program that runs \\
\code{FuzzDebugfsQuote} & image & a quoted path never breaks out of its line or quotes in a \code{debugfs} command file \\
\code{FuzzFixups} & image & every \code{debugfs} fix-up addresses a file that exists in the staged tree \\
\code{FuzzParse} & config & any \code{tkbox.yaml} fails or yields a config whose limits are in range, whose secrets have one source and exact hosts, and which validates again \\
\code{FuzzParseSecretFlag} & config & \code{--secret} parses only to a valid secret; a refused flag's error never shows what may be a pasted value \\
\bottomrule
\end{tabularx}
\end{table*}

\paragraph{Findings.} Fuzzing produced nine findings, all fixed (\cref{tab:findings}): eight from the campaign that wrote the targets, and one (B2) that \code{FuzzRedactReader} found later in the repository's own \code{make fuzz} run. Six are in the egress proxy, where the guarantees on secrets and addresses depend on them. F3 is the most serious: Go's \code{net/http} parses \code{GET https:http://evil.example/x} as an opaque URL with no host, so the inner-request check passed it, and the proxy then wrote it upstream as an absolute-form request for \code{evil.example} with the real secret swapped in. A front end that routes by absolute-form target could have handed the credential to another site while the record named only the bound host. F2 let a response with a status below 100 relay its body unredacted, so a bound host echoing the swapped value could have shown it to the guest. F4 allowed a redaction marker to combine with neighbouring bytes into the value it replaced, and B2 let a body read in pieces keep parts of a secret whose value contains another secret's.

\begin{table*}[t]
\centering\small
\caption{Fuzzing findings, all fixed, with the labels used in the repository's tests and commits.}\label{tab:findings}
\begin{tabularx}{\textwidth}{@{}P{0.6cm}P{1.5cm}Y@{}}
\toprule
ID & Area & Finding and fix \\
\midrule
F1 & patch parser & \code{/dev/null} on a \code{---} or \code{+++} line without its \code{new file mode} or \code{deleted file mode} line is a file named \code{dev/null} to git, not a creation or deletion; now refused \\
F2 & egress & a response status below 100 had its body relayed without redaction; now a 502 \\
F3 & egress & an opaque request target inside a TLS tunnel could carry a swapped secret to another host; now refused \\
F4 & egress & a redaction marker could rebuild a value with the bytes around it; markers can no longer take part in a match \\
F5 & commands & a program name starting with \code{\%} is a job spec to bash and zsh, so the decided command line would not name what runs; now refused \\
F6 & config & a refused \code{--secret} flag could quote part of a pasted value in its error; the source is now checked first and never quoted \\
I1 & egress & an IPv6 \code{--egress-block} prefix spanning the IPv4-mapped range did not refuse the IPv4 addresses it covers \\
I2 & egress & SRv6 segment IDs (\code{5f00::/16}) were missing from the refused ranges \\
B2 & egress & the streaming redactor, reading a body in pieces, could leave parts of a secret whose value contains another secret's; found by \code{FuzzRedactReader} after the campaign \\
\bottomrule
\end{tabularx}
\end{table*}

\subsection{Review}\label{sec:review}

Before release the code went through rounds of independent review, followed by verification passes that checked each fix and each claim in the documentation against the code. The README counts three independent review rounds plus the fuzzing campaign; the commit history labels fixes from each pass (\cref{tab:review}). The findings were of the same kind as the fuzz findings: ways a hostile guest, a hostile upstream or a hostile patch could make a door do something other than what the record says.

\begin{table}[t]
\centering\footnotesize
\caption{Review passes, as labelled in the commit history, with examples of what each fixed.}\label{tab:review}
\begin{tabularx}{\linewidth}{@{}P{1.45cm}Y@{}}
\toprule
Pass & Examples of fixes \\
\midrule
review-1 & a \code{CONNECT} to 443 under an entry naming 443 recorded as \code{tcp://}; base64 values that are JSON-escaped or percent-encoded redacted \\
re-review & gaps in apply, stdin, the observer, events and MCP limits \\
verify-3, 3b & documentation claims corrected; apply stated plainly to have no rollback \\
verify-5 & more per-exec env names that run code refused; apply writes into a global \code{core.hooksPath} refused; an approval lost after a stopping box wrote its gap record now recorded \\
final checks & a run not closed under an approval record still in flight; no part of a nested secret value shown in a refusal or event; three guarantees backed only by code given tests \\
\bottomrule
\end{tabularx}
\end{table}

\subsection{Summary}

\Cref{tab:evalsum} collects the numbers this section reports.

\begin{table}[t]
\centering\footnotesize
\caption{The evaluation at a glance.}\label{tab:evalsum}
\begin{tabularx}{\linewidth}{@{}Yr@{}}
\toprule
Item & Result \\
\midrule
End-to-end checks on KVM (CI run \#16) & 13 / 13 pass \\
\code{TestBox} (the 13 checks), including pull and boot & 28.5\,s \\
Unit tests; end-to-end test functions & 340; 2 \\
Fuzz targets; findings fixed & 15; 9 \\
Demo bundle: records; decisions; verifies & 102; 46; yes \\
Assurance of every bundle reported & dev \\
\bottomrule
\end{tabularx}
\end{table}

\section{Discussion}\label{sec:discussion}

\subsection{Limits}\label{sec:limits}

Box v0 is a first release. Its limits fall into three groups: what the design does not try to do, what the implementation does not do yet, and what the evaluation has not shown.

\paragraph{By design.}
\begin{itemize}[leftmargin=*]
  \item \textbf{Inside the VM is not recorded.} Box records doors, not processes. A command started from the host is decided; the processes it starts are not, and an agent that runs a script which runs a hundred commands leaves one \code{Bash} decision and whatever those commands did at the doors. Box v0 runs no process sensor in the guest.
  \item \textbf{Who, not what.} Box decides which hosts the agent may reach, not what it sends them. Any allowed host that fetches URLs for its callers is a relay: \code{go-proxy} fetches modules from whatever host a module path names, so an agent can send workspace data to a domain of its choosing through an allowed host. The documentation says so, and the only fix is a smaller allowlist.
  \item \textbf{Uninspected connections.} With the default \code{inspect: bound}, the proxy reads only connections to hosts with a bound secret, and plain HTTP. On other HTTPS hosts it sees the host, port and server name. A secret's \emph{value} cannot leave there, because it is not in the VM; its placeholder can, and is useless to the receiver. \code{inspect: all} reads every allowed HTTPS host, at the cost of breaking clients that pin certificates.
  \item \textbf{Only \code{/workspace} comes back.} Changes elsewhere in the VM are lost at stop.
  \item \textbf{No rollback.} \code{tkbox apply} decides every file before it writes any, and re-checks the checkout before writing, but if a write fails midway it names exactly which paths were written and does not undo them.
\end{itemize}

\paragraph{Not in v0.} \Cref{tab:notv0} lists what is not built yet.

\begin{table}[t]
\centering\footnotesize
\caption{Not in v0, and what happens today.}\label{tab:notv0}
\begin{tabularx}{\linewidth}{@{}P{2.2cm}Y@{}}
\toprule
Not in v0 & Today \\
\midrule
Firecracker jailer & Firecracker runs as the daemon's user with its default seccomp filter, without the jailer's chroot, namespaces and cgroups; the VMM interface already takes a jailer config \\
Remote signer & the signer must be on the daemon's host (Unix socket only) \\
Private registries & public images only \\
Snapshots and restore & each box boots from its image; a stopped box cannot be resumed \\
Kubernetes RuntimeClass & run \code{tkbox daemon} on the node or runner directly \\
gVisor fallback & no \code{/dev/kvm}, no daemon \\
macOS hosts & the CLI builds on macOS; boxes need Linux (a Lima VM works) \\
SSH and UDP egress & HTTPS and HTTP only; no DNS, QUIC or NTP from the box \\
GPUs, Windows guests & none; Linux images only \\
\bottomrule
\end{tabularx}
\end{table}

\paragraph{Operational caveats.}
\begin{itemize}[leftmargin=*]
  \item \textbf{The host's public address on a cloud VM.} Box refuses the host's own addresses as it sees them on its interfaces. On EC2, GCE, Azure and GitHub-hosted runners the public IPv4 is translated and on no interface, so a box allowed to reach it could reach services the host listens on, unless the operator passes it with \code{--egress-block}.
  \item \textbf{Same-user access.} The daemon's API accepts any process of its own user. An MCP orchestrator that can also run a shell as that user can bypass MCP's narrowing and approve its own connections. Operators should run the orchestrator as another user.
  \item \textbf{Containment latency.} A hold or quarantine set on the signer takes effect within 2\,s in a box's first minute, and within about 15\,s after that, because the daemon polls.
  \item \textbf{ECH and HTTP/2.} Clients that send an encrypted ClientHello are refused, since Box must see the server name; Chromium-based tools may hit this. Inspected connections speak HTTP/1.1 only.
\end{itemize}

\paragraph{What the evaluation has not shown.} As \cref{sec:eval} says: no performance numbers beyond the suite's wall-clock times, no study with a capable agent trying to get out, and no deployment above dev assurance.

\subsection{Why doors, and not a smarter guest}

An alternative design would put monitoring inside the guest: a syscall tracer, an eBPF sensor, a hardened agent runtime. We chose the opposite because everything inside the guest is the attacker's. A monitor in the guest can be killed, starved or lied to by code running as root next to it, and its records would have to be trusted on the guest's word. A door on the host cannot be bypassed from inside, because there is no other way out of a VM with no network card, and the record of what passed the door is made by the host, not reported by the guest. The cost is the first limit above: Box sees only what crosses a door. We think this is the right trade for the actions that cause harm outside the box, which are exactly the ones that must cross a door to cause it.

\subsection{Future work}

\begin{itemize}[leftmargin=*]
  \item \textbf{Jailer and stronger host hardening.} Running Firecracker under its jailer (chroot, namespaces, cgroups, a per-box user) is the most important missing piece of defence in depth.
  \item \textbf{Remote signer and witnesses.} A signer on another host, reached over an authenticated channel, and a witness that cosigns checkpoints, so that a bundle proves more than dev assurance.
  \item \textbf{An adversarial agent study.} Put capable agents in boxes, with instructions or injected text that tell them to get data or a key out, and report what they try and where it is stopped.
  \item \textbf{Performance.} Boot latency, exec overhead, proxy throughput, and the cost of a signer decision per connection, against containers and other micro-VM sandboxes.
  \item \textbf{Processes inside the VM.} A guest sensor whose records are reconciled against the doors, so that a gap between what the guest says it ran and what reached a door is visible.
  \item \textbf{Deployment.} Private registries, snapshots for fast start, and a Kubernetes RuntimeClass.
  \item \textbf{More Verified Runs.} One for every agent family, so the table measures how real agents behave inside a box.
\end{itemize}

\section{Conclusion}\label{sec:conclusion}

Tracekit Box runs an AI agent in a Firecracker micro-VM whose only ways out are doors on the host, and puts a Tracekit signer at every door. Commands, connections and file writes are decided before they happen; secrets stay on the host and are used on the agent's behalf only in one header for exact hosts; the agent's work comes back as a patch applied file by file; and every decision, use and refusal is signed into one run per box that anyone can verify offline. The v0 implementation passes all of its end-to-end checks on real KVM hardware, its trust-boundary parsers are fuzzed against stated properties, and the issues that fuzzing and review found are fixed. Its limits are stated plainly: it records doors, not processes inside the VM; it decides who the agent may reach, not what it says; it does not yet run the jailer; and every bundle reported here has dev assurance. Within those limits, it gives an operator something sandboxes and tracing do not give separately: a box the agent cannot leave unseen, and a record of what left it that the operator does not have to take on trust.

\section*{Acknowledgements}
We thank the Firecracker maintainers, whose virtual machine monitor Box builds on, and the authors of the C2SP and Certificate Transparency specifications behind Tracekit's evidence format. Any errors are ours.

\footnotesize
\setlength{\bibsep}{2pt plus 1pt}
\bibliographystyle{abbrvnat}
\bibliography{references}

@inproceedings{greshake2023indirect,
  title     = {Not What You've Signed Up For: Compromising Real-World {LLM}-Integrated Applications with Indirect Prompt Injection},
  author    = {Greshake, Kai and Abdelnabi, Sahar and Mishra, Shailesh and Endres, Christoph and Holz, Thorsten and Fritz, Mario},
  booktitle = {Proceedings of the 16th ACM Workshop on Artificial Intelligence and Security (AISec)},
  year      = {2023},
  note      = {arXiv:2302.12173}
}

@inproceedings{zhan2024injecagent,
  title     = {{InjecAgent}: Benchmarking Indirect Prompt Injections in Tool-Integrated Large Language Model Agents},
  author    = {Zhan, Qiusi and Liang, Zhixiang and Ying, Zifan and Kang, Daniel},
  booktitle = {Findings of the Association for Computational Linguistics: ACL 2024},
  year      = {2024},
  note      = {arXiv:2403.02691}
}

@inproceedings{debenedetti2024agentdojo,
  title     = {{AgentDojo}: A Dynamic Environment to Evaluate Prompt Injection Attacks and Defenses for {LLM} Agents},
  author    = {Debenedetti, Edoardo and Zhang, Jie and Balunovi{\'c}, Mislav and Beurer-Kellner, Luca and Fischer, Marc and Tram{\`e}r, Florian},
  booktitle = {Advances in Neural Information Processing Systems, Datasets and Benchmarks Track},
  year      = {2024},
  note      = {arXiv:2406.13352}
}

@misc{rfc6962,
  author       = {Ben Laurie and Adam Langley and Emilia Kasper},
  title        = {Certificate Transparency},
  howpublished = {RFC 6962},
  year         = {2013},
  url          = {https://www.rfc-editor.org/rfc/rfc6962}
}

@misc{tracekit,
  author       = {Bravish Ghosh},
  title        = {{Tracekit: Tamper-Evident Intent--Reasoning--Action Auditing for AI Agents}},
  howpublished = {arXiv:2609.35659},
  year         = {2026},
  note         = {Code: \url{https://github.com/Cygnux-Labs/Tracekit}}
}

@misc{openai_dns,
  author       = {{OpenAI}},
  title        = {An agent used {DNS} to reach an external chatbot},
  howpublished = {OpenAI Alignment blog, misalignment reports},
  year         = {2026},
  url          = {https://alignment.openai.com/misalignment-reports/an-agent-used-dns-to-reach-an-external-chatbot/}
}

@inproceedings{agache2020firecracker,
  title     = {Firecracker: Lightweight Virtualization for Serverless Applications},
  author    = {Agache, Alexandru and Brooker, Marc and Florescu, Andreea and Iordache, Alexandra and Liguori, Anthony and Neugebauer, Rolf and Piwonka, Phil and Popa, Diana-Maria},
  booktitle = {17th USENIX Symposium on Networked Systems Design and Implementation (NSDI)},
  year      = {2020}
}

@inproceedings{young2019gvisor,
  title     = {The True Cost of Containing: A {gVisor} Case Study},
  author    = {Young, Ethan G. and Zhu, Pengfei and Caraza-Harter, Tyler and Arpaci-Dusseau, Andrea C. and Arpaci-Dusseau, Remzi H.},
  booktitle = {11th USENIX Workshop on Hot Topics in Cloud Computing (HotCloud)},
  year      = {2019}
}

@misc{kata,
  author       = {{Kata Containers project}},
  title        = {{Kata Containers}},
  howpublished = {\url{https://katacontainers.io}},
  year         = {2026}
}

@misc{e2b,
  author       = {{E2B}},
  title        = {{E2B}: Open-source cloud sandboxes for {AI} agents},
  howpublished = {\url{https://github.com/e2b-dev/E2B}},
  year         = {2026}
}

@misc{microsandbox,
  author       = {{microsandbox contributors}},
  title        = {microsandbox: self-hosted micro-{VM} sandboxes for untrusted code},
  howpublished = {\url{https://github.com/superradcompany/microsandbox}},
  year         = {2026}
}

@misc{srt,
  author       = {{Anthropic}},
  title        = {sandbox-runtime: operating-system sandboxing with network filtering for agents},
  howpublished = {\url{https://github.com/anthropic-experimental/sandbox-runtime}},
  year         = {2026}
}

@misc{openshell,
  author       = {{NVIDIA}},
  title        = {{OpenShell}: a sandbox runtime for {AI} agents},
  howpublished = {\url{https://github.com/NVIDIA/OpenShell}},
  year         = {2026}
}
\normalsize

\onecolumn
\raggedbottom
\appendix
\section{Quick look}\label{app:quick}

On a Linux host with KVM, with Tracekit 1.4.2 installed:

\begin{lstlisting}
pip install tracekit-ai==1.4.2
tracekit init --dev --v2 --no-hooks && tracekit up --wait   # a same-user dev signer
tkbox daemon &                                 # refuses to start without the signer
tkbox run --image python:3.12-slim --allow pypi,debian --workspace . --name job \
          --secret GH=env:GH_TOKEN@api.github.com
tkbox exec job -- sh -c 'apt-get update -q && apt-get install -y -q curl'
                                               # the image has no curl: debian preset
tkbox exec job -- pip install requests         # allowed by the pypi preset
tkbox exec job -- curl -sI https://example.com # held: answer with `tkbox approve <id>`
tkbox exec job -- sh -c 'echo $GH'             # tkbox_GH_<32 hex>, never the token
tkbox stop job                                 # saves the patch
tkbox diff job                                 # review it
tkbox apply job                                # file by file, each decided by the signer
tkbox export job -o run.tkb                    # run.tkb + run.trust.json
tracekit verify run.tkb --trust run.trust.json # Integrity: VERIFIED.
\end{lstlisting}

\section{Example configuration}\label{app:config}

\code{tkbox.yaml} at a repository root, abridged from \code{examples/tkbox.yaml}. Flags given to \code{tkbox run} override these values.

\begin{lstlisting}
image: python:3.12-slim
name: service-agent
workspace: .            # copied into /workspace: never .git, never credential files
allow: [pypi, github, anthropic]   # anything else waits 2 minutes for `tkbox approve`
secrets:
  ANTHROPIC_API_KEY:
    from_env: ANTHROPIC_API_KEY
    hosts: [api.anthropic.com]     # exact names, no wildcards
    header: X-Api-Key
  GH_TOKEN:
    from_env: GH_TOKEN
    hosts: [api.github.com, github.com]
    header: Authorization
env:                    # visible in the box and recorded: never secrets
  PIP_DISABLE_PIP_VERSION_CHECK: "1"
limits:
  cpus: 2
  memory: 4GiB          # writes outside /workspace use this memory
  disk: 8GiB            # for writes to /workspace
  timeout: 1h           # the box's lifetime
  exec_timeout: 30m     # each command's
inspect: bound          # all: read every allowed HTTPS host; refuse unlisted hosts at once
\end{lstlisting}

\section{Default limits}\label{app:limits}

\begin{table}[H]
\centering\small
\caption{Selected defaults (the full list, with each message Box prints, is in the repository's \code{docs/limits.md}).}\label{tab:defaults}
\begin{tabularx}{\linewidth}{@{}Yr@{}}
\toprule
Limit & Default \\
\midrule
vCPUs; memory; workspace disk & 2; 2\,GiB; 4\,GiB \\
Box lifetime; each command & 1\,h; 30\,min \\
Boxes per daemon; concurrent execs per box & 8; 64 \\
Open connections per box & 256 \\
Pending approvals per box; new approvals per minute & 16; 30 \\
Approval wait & 2\,min \\
Connect, resolve and TLS handshake timeouts & 10\,s each \\
Idle connection & 5\,min \\
Request head; upstream response head & 64\,KiB; 1\,MiB \\
Up-front stdin; decided as text if UTF-8 and at most & 1\,MiB; 4\,KiB \\
Secret value size & 64\,KiB, one line \\
Patch size & 64\,MiB \\
Workspace size & 100,000 files, 4\,GiB \\
Must-keep observation queue & 4096 \\
Box's own refusal records & 20/s per box \\
Containment poll & 2\,s (first minute), then 15\,s \\
\bottomrule
\end{tabularx}
\end{table}

\section{Glossary}\label{app:glossary}

\begin{description}[leftmargin=0pt,itemsep=2pt]
  \item[Box] One Firecracker micro-VM running one agent job, with its own egress proxy, CA and Tracekit run.
  \item[Door] A channel out of the box that the host serves and the signer decides: commands, network, secrets, model calls, changes, stop.
  \item[Decision] A signer evaluation of a tool call (\code{Bash}, \code{WebFetch}, \code{Write}) before it happens: allow, flag, ask or deny.
  \item[Observation] A signed record the signer does not decide, such as \code{tkbox.secret.use}.
  \item[Placeholder] The value \code{tkbox\_<NAME>\_<32 hex>} that the VM sees in place of a secret.
  \item[Bound host, bound header] The exact hosts, and the one header, where the proxy swaps a placeholder for the real value.
  \item[Approval] A person on the host allowing or rejecting a connection to a host off the allowlist with \code{tkbox approve} or \code{tkbox reject}.
  \item[Gap] A \code{tkbox.gap} observation that counts records Box had to keep and could not.
  \item[Bundle] The \code{.tkb} file \code{tkbox export} writes: a box's run, its checkpoint and its keys, verifiable offline with its trust file.
  \item[Assurance] What the verifier says about who could have signed the records: \code{dev} for a same-user dev signer.
\end{description}

\end{document}